\documentclass[12pt,reqno]{article}
\usepackage{setspace,graphicx,amssymb,amsmath,latexsym,amsfonts,amscd,amsthm,multirow,ctable,mathdots,caption,array,diagbox,mathtools}
\usepackage{chet}
\usepackage{multicol}
\usepackage{tabularx,cite,mathrsfs}
\usepackage{authblk}
\usepackage{color}

\newcommand{\es}[2] {\begin{equation} \label{#1} \begin{split} #2 \end{split} \end{equation}}

\usepackage{fullpage}
\usepackage{stmaryrd}
\usepackage{rotating}

\usepackage{hyperref}

\usepackage{bbm}

\newcommand{\be}{\begin{eqnarray}}
\newcommand{\ee}{\end{eqnarray}}
\newcommand{\eeq}{\end{equation}}
\newcommand{\beq}{\begin{equation}}

\usepackage{simplewick}
\def\OO{{\cal O}}

\renewcommand{\b}{\bar}
\allowdisplaybreaks \numberwithin{equation}{section}
\DeclareSymbolFont{AMSa}{U}{msa}{m}{n}
\DeclareSymbolFont{AMSb}{U}{msb}{m}{n}
\DeclareMathSymbol{\fieldR}{\mathalpha}{AMSb}{"52}

\newcommand{\CF}{{\cal F}}

\newcommand{\CM}{{\cal M}}

\newcommand{\CO}{{\cal O}}
\newcommand{\CP}{{\cal P}}

\newcommand{\tr}{{\rm tr}}

\def\beq{\begin{equation}}
\def\eeq{\end{equation}}
\def\bea{\begin{eqnarray}}
\def\eea{\end{eqnarray}}

\def\<{\left\langle}

\newcommand\nn{\nonumber}
\newcommand{\abs}[1]{\left| #1 \right|}

\def\X{\mathcal{X}}
\newcommand{\rep}[4]{ \rho \left(
    \begin{array}{cc}
      #1 & #2\\
      #3 & #4
    \end{array}
  \right)}

\newcommand{\sltz}{SL(2,\mathbb{Z})}

\renewcommand{\r}{\rho}

\renewcommand{\b}{\beta}
\newcommand{\n}{\nu}
\newcommand{\m}{\mu}

\renewcommand{\(}{\left(}
\renewcommand{\)}{\right)}
\renewcommand{\[}{\left[}
\renewcommand{\]}{\right]}

\newcommand{\cali}{\includegraphics[height=3.725mm, trim = 0mm 0mm 0mm 0mm, clip]{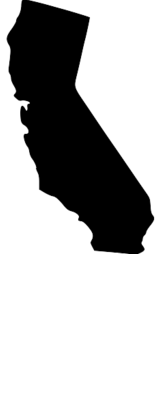}}
\newcommand{\mass}{\includegraphics[height=3.725mm, trim = 0mm 0mm 0mm 0mm, clip]{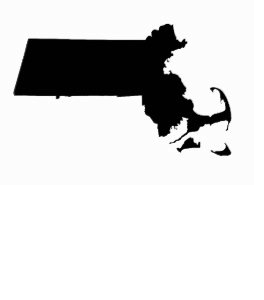}}
\newcommand{\nj}{\includegraphics[height=3.725mm, trim = 0mm 0mm 0mm 0mm, clip]{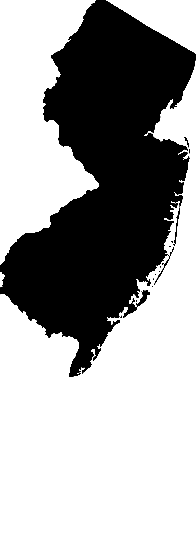}}
\newcommand{\maryland}{\includegraphics[height=3.725mm, trim = 0mm 0mm 0mm 0mm, clip]{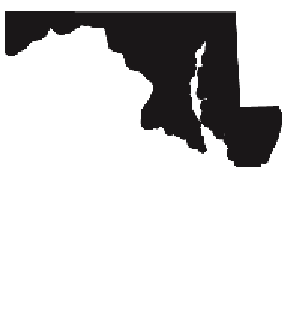}}
\newcommand{\goog}{\includegraphics[height=3.725mm, trim = 0mm 0mm 0mm 0mm, clip]{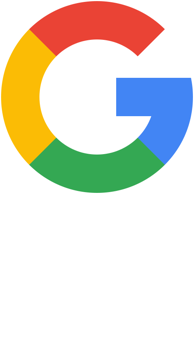}}

\begin{document}

\setstretch{1.2}

\begin{center}

{
~ \\ ~ \\ ~\\ ~\\ ~\\ ~\\~
\Huge The Most Irrational Rational Theories}
\end{center}
\vspace{0.5cm}
\begin{center}
         {Nathan Benjamin\cali~\nj,}
           {Ethan Dyer\cali~\maryland~\goog},
                  {A. Liam Fitzpatrick\mass},
          {Yuan Xin\mass}\\

\vspace{0.5cm}\cali {\it Stanford Institute for Theoretical Physics, Via Pueblo, Stanford, CA 94305, USA}\\
\mass {\it Boston University Physics Department, Commonwealth Avenue, Boston, MA 02215, USA}\\
\nj {\it Princeton Center for Theoretical Science, Princeton University, Princeton, NJ 08544, USA}\\
\maryland {\it Department of Physics and Astronomy, Johns Hopkins University, Charles Street, Baltimore, MD 21218, USA}\\
\goog {\it Google, Amphitheatre Parkway, Mountain View, CA 94043, USA}

\end{center}

\begin{center}
  {\bf Abstract}
\end{center}

We propose a two-parameter family of modular invariant partition functions of two-dimensional conformal field theories (CFTs) holographically dual to pure three-dimensional gravity in anti de Sitter space. Our two parameters control the central charge, and the representation of $SL(2,\mathbb{Z})$. At large central charge, the partition function has a gap to the first nontrivial primary state of $\frac{c}{24}$.  As the $SL(2,\mathbb{Z})$ representation dimension gets large, the partition function exhibits some of the qualitative features of an irrational CFT. This, for instance, is captured in the behavior of the spectral form factor. As part of these analyses, we find similar behavior in the minimal model spectral form factor as $c$ approaches $1$.

\clearpage

\tableofcontents

\newpage
\section{Introduction}

Since the advent of the AdS/CFT correspondence \cite{Maldacena:1997re}, progress on quantum gravity and on understanding and constructing large families of CFTs have been inextricably linked.  A complete solution of a CFT with a large radius gravity dual would likely shed light on many subtle questions about bulk gravitational physics.  However, there is a general tension between simplicity  of a CFT and proximity of its gravity dual to Einstein gravity.  
It is well-understood that a sparse light field theory spectrum is a necessary ingredient for a local weakly curved gravitational dual \cite{Heemskerk:2009pn}. In three dimensions the gravitational phase structure imposes a sharp constraint on the density of light states \cite{Hartman:2014oaa}. A spartan approach to this constraint is to look for theories dual to pure gravity, i.e. to attempt to build theories with as large a gap between the vacuum and the next state as possible.

Witten explored this question for chiral or holomorphically factorized conformal field theories \cite{Witten:2007kt}. In this case, the gap to the lightest non vacuum state is bounded by $c/24+1$. Candidate theories with this gap have a unique partition function, and are parameterized by single integer, $k = c/24$, controlling the central charge. For $k=1$ this theory is the monster CFT\cite{FLM}; for higher $k$, no theories are known and various no go theorems have been established \cite{Witten:2007kt, Gaberdiel:2007ve, Gaiotto:2008jt, Gaberdiel:2008pr}. Perhaps most strikingly, for higher amounts of supersymmetry, one can explicitly rule out these maximally gapped, extremal, theories \cite{Gaberdiel:2008xb}. These no go theorems typically rely on the harshest form of extremality, and can be naturally evaded by the addition of a few lighter states \cite{Benjamin:2016aww}. A more conceptual criticism of these theories, mentioned even in the original work, is that, as a result of their chiral nature, they are essentially integrable. All states have integer dimension, and there exist a tower of higher spin conserved currents. It is becoming increasingly clear that this is not a feature of typical Einstein-like gravity theories, which exhibit chaotic type behavior for many observables\cite{Shenker:2013pqa, Shenker:2013yza, Maldacena:2015waa, Cotler:2016fpe}.

Going beyond chiral constructions, modular bootstrap techniques allow one to put constraints on the gap to the first excited states in any 2d CFT \cite{Hellerman:2009bu, Friedan:2013cba, Collier:2016cls}. The numerical current state of the art is that the bound to the first excited state in any 2d CFT must be at most $\frac{c}8 + \frac12$ above the vacuum for $c>4$. Interestingly, integrable theories rear their heads again in the modular bootstrap approach. When looking at bounds in the gap to the first excited scalar in 2d CFT, \cite{Collier:2016cls} (see also \cite{Bae:2017kcl,Dyer:2017rul}) noticed that, at small $c$, there are kinks saturated by the $SU(2)$ WZW model at level 1 ($c=1$), the $SU(3)$ WZW model at level 1 ($c=2$), the $G_2$ WZW model at level 1 ($c=\frac{14}5$), and the $SO(8)$ WZW model at level 1 ($c=4$).

There appears to be a dichotomy. Integrable constructions easily give candidate large gap partition functions, and at small central charge there do exist both chiral and non-chiral integrable large gap theories, yet we know this integrality is a feature we'd like to do away with for Einstein-like gravity.
In this paper we try to take the best of both worlds.
We borrow technology originally introduced by Bantay and Gannon to describe the modular properties of characters of rational conformal field theories \cite{Bantay:2005vk}.
These \textit{vector valued modular forms} (VVMFs) transform under finite-dimensional representations of the modular group $\sltz$, and we use this to build candidate partition functions with sparse light spectra.
We present constructions of large gap candidate partition functions with positive integral spectrum. The degree of integrability is related to the representation dimension, and can be tuned. These functions represent the first examples of truly non-holomorphically-factorized large gap candidate partition functions\footnote{Strictly speaking, one could take linear combinations of large gap holomorphically factorized functions of the form in \cite{Witten:2007kt}, but each term added would lower the gap, and moreover each term would be individually modular invariant unlike in our case.}.

As we take the representation large, the partition functions exhibit more irrational-like behavior; for instance the recurrence time becomes arbitrarily large\footnote{The authors of \cite{Balasubramanian:2016ids} had a similar motivation in the study of the D1/D5 system at the orbifold.}.
Another sharp feature is obtained by looking at the spectral form factor (SFF), an analytic continuation of the thermal partition function.
At large dimension, we show evidence that the SFF exhibits the dip-ramp-plateau structure characteristic of chaotic, gravitational theories \cite{Cotler:2016fpe}.
\*

The paper is organized as follows. In Section \ref{sec:minmod}, we review the spectral form factor and analyze it for minimal models at large $m$. In particular, we show that for large $m$ the minimal model SFF exhibits a dip, linearly growing ramp, and plateau, reminiscent of chaotic theories. In Section \ref{sec:modular}, we describe our algorithm to go to large central charge, while preserving the $\sltz$ structure and keeping integrality and positivity. In Section \ref{sec:largegapsff}, we analyze the spectral form factor for our partition functions at large $c$ and large $m$. Finally in Section \ref{sec:discuss}, we discuss potentially interesting questions. Some detailed calculations are relegated to the appendices. Readers interested in the large $c$ construction can skip to Section \ref{sec:modular}; readers interested the analysis of the spectral form factor for our candidate theories of pure gravity can skip to Section \ref{sec:largegapsff}; readers solely interested in whether or not we cited them can skip to page \pageref{refs}.

\section{Minimal Models at Large \texorpdfstring{$m$}{m} }
\label{sec:minmod}

In practically all areas of physics, solvable toy models play a crucial role in building up physical intuition and providing guidance about what kinds of physical effects are possible.  In the case of quantum gravity, however, ergodicity is a key characteristic of the physics one would like to understand, creating an obvious obstacle for solvable models.  A concrete manifestation of this tension is that in a representative toy model of gravity, time evolution of states should typically take a parametrically long time before passing back close to the initial state, i.e. the recurrence time should be long.  However, in rational models, the energies are all rational numbers with a finite least common denominator $d$,  and so time evolution is periodic with period (at most) $d$.

A useful quantitative measure of the ergodicity of gravity theories is provided by the Spectral Form Factor (SFF) $g(\beta,t)$, defined in terms of the partition function $Z(\beta)$ as
\be
\label{sffdef}
g(\beta,t) \equiv |Z(\beta+i t)|^2 = \sum_{i,j} e^{-\beta (E_i+ E_j) + i t (E_i - E_j)},
\ee
where the sum on $i,j$ is over all states in the theory. The SFF encodes a concrete formulation of the information paradox: in a semiclassical gravity treatment,  $g(\beta,t)$ decays exponentially to zero at late times, whereas in a unitary theory with a discrete spectrum its late-time average must have a positive lower bound (see e.g. \cite{Dyer:2016pou} for a simple proof).  This tension can be understood purely in CFT language as well, by studying the behavior of individual Virasoro characters at large central charge $c$ \cite{Dyer:2016pou}.   Moreover, it has been proposed \cite{Cotler:2016fpe} that at intermediate times, the SFF should exhibit additional features characteristic of random matrix theory.

A useful toy model for the resolution of this version of the information paradox should satisfy the microscopic conditions of unitarity and a discrete spectrum while simultaneously having a limit where semiclassical gravity behavior emerges.  The extremal chiral partition functions considered by Witten \cite{Witten:2007kt} satisfy the large $c$, large gap assumptions required for gravity as well as the microscopic conditions of unitarity and a discrete spectrum. Consequently, these partition functions provide concrete examples that exhibit the ``semiclassical'' early time decay as well as the ``unitary'' late-time lower-bound.  From a gravity path integral point of view, this is fairly non-trivial, and involves summing over an infinite number of gravitational saddle points using the method of Rademacher sums \cite{Maloney:2007ud}.  However, from a Hamiltonian point of view, the late-time lower-bound is trivial.  The reason is that in chiral theories, the (radial quantization) energy $\Delta$ of any state is the same as its spin $L$ and is therefore an integer, so time evolution of any state is periodic with period (at most) $2\pi$.\footnote{Some examples also contain fermions, in which case the period is at most $4\pi$.} So the SFF also has a very short period, $g(\beta, t)=g(\beta,t+2\pi)$, which does not leave enough room for the more distinctive features expected at intermediate times. This behavior reflects the fact that in chiral theories, every state is associated with a conserved quantity.  Unsurprisingly, such theories are not particularly chaotic.

To improve on the chiral extremal partition functions, we would therefore like to generalize the construction to nonchiral theories containing states with incommensurable energies, i.e. $E_i/E_j$ is irrational.  As discussed above, we will settle for a weaker condition that the least common denominator $d$ taken among all states in the theory is large, so that the period of the SFF is very long.  Ultimately, we are interested in making this generalization while simultaneously maintaining a large central charge and sparse spectrum.  In this section, however, we will begin by exploring the minimal models at large $m$ as unitary examples with a large common denominator $d \sim \CO(m^2)$ and by analyzing the features of their SFFs.  In later sections, we will discuss how to combine elements of the minimal models and Witten's extremal examples to get theories with large $c$ and large gap, using the methods of \cite{Bantay:2005vk}. We pause here to emphasize that any infinite sequence of RCFTs where the least common denominator $d$ amongst states becomes arbitrarily large could in principle exhibit similar features; in this paper, we focus on the Virasoro minimal models simply as an explicit example. It would be interesting to repeat our analysis for other families of RCFTs, for instance some family of WZW models.

\subsection{SFF}
\label{sec:minimal_models}
The $\tau = -\bar\tau$ projected partition function of the diagonal minimal model $\mathcal{M}(m, m+1)$ is 
\be
Z_m(\tau) &=& \frac{Z_m^0(\tau) }{2 \eta(\tau)^2 },
\ee
where
\begin{align}
  Z_m^0(\tau) \equiv \vartheta _3\left(0,e^{\frac{i \pi  \tau }{m (m+1)}}\right)
  \vartheta _3\left(0,e^{i m (m+1) \pi \tau }\right)
  -\vartheta _3\left(0,e^{\frac{i m \pi  \tau }{m+1}}\right)
  \vartheta_3\left(0,e^{\frac{i (m+1) \pi  \tau }{m}}\right)~,
  \label{eq:MinModelZ}
\end{align}
and our convention for the Jacobi $\vartheta_3$ function is  $\vartheta _3\left( w,q \right):= \sum_{n=-\infty}^{\infty} (w^2)^n q^{n^2}$.
We derive this relation for $Z_m$ in Appendix \ref{app:minmodelZ}.

Before exploring the behavior of the minimal model SFFs in detail, let us make a few general comments. In random matrix theories, the early-time decay dips below the late-time plateau, and approaches the plateau with a ramp-like behavior before flattening out (we give a brief review in section \ref{sec:largegapsff}). As mentioned previously, the minimal models have a common denominator $2m(m+1)$ and therefore their SFF always contains an overall recurrence $g(\beta+i t_{\rm rec}) = g(\beta)$ with
\be
t_{\rm rec} = 4 \pi m(m+1).
\ee
The minimal model index $m$ must be taken large in order for dip and ramp features to have time to emerge before the recurrence time $t_{\rm rec}$.
\begin{figure}[htbp]
  \centering
  \includegraphics[width=0.45\linewidth]{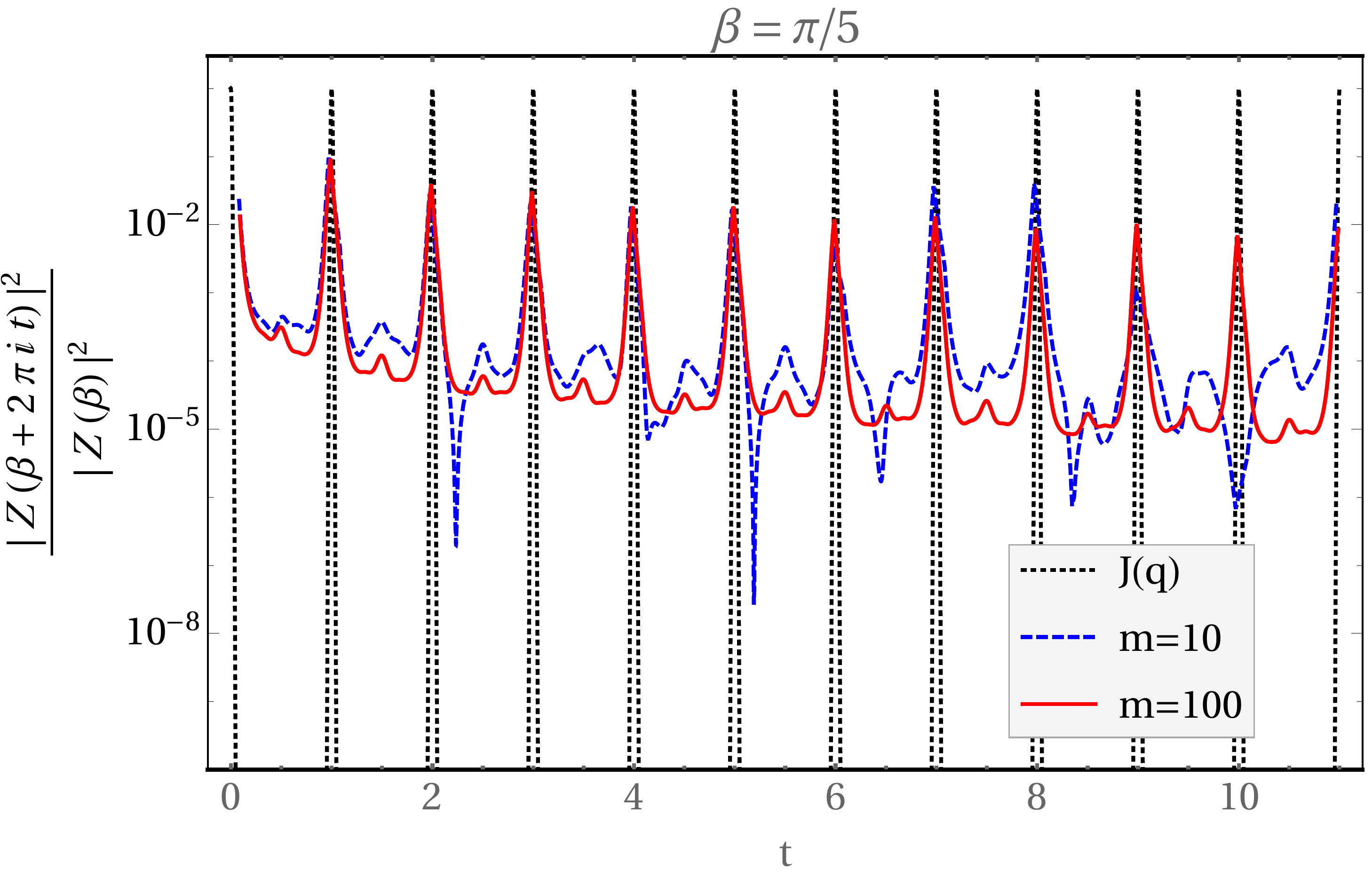}
  ~\includegraphics[width=0.45\linewidth]{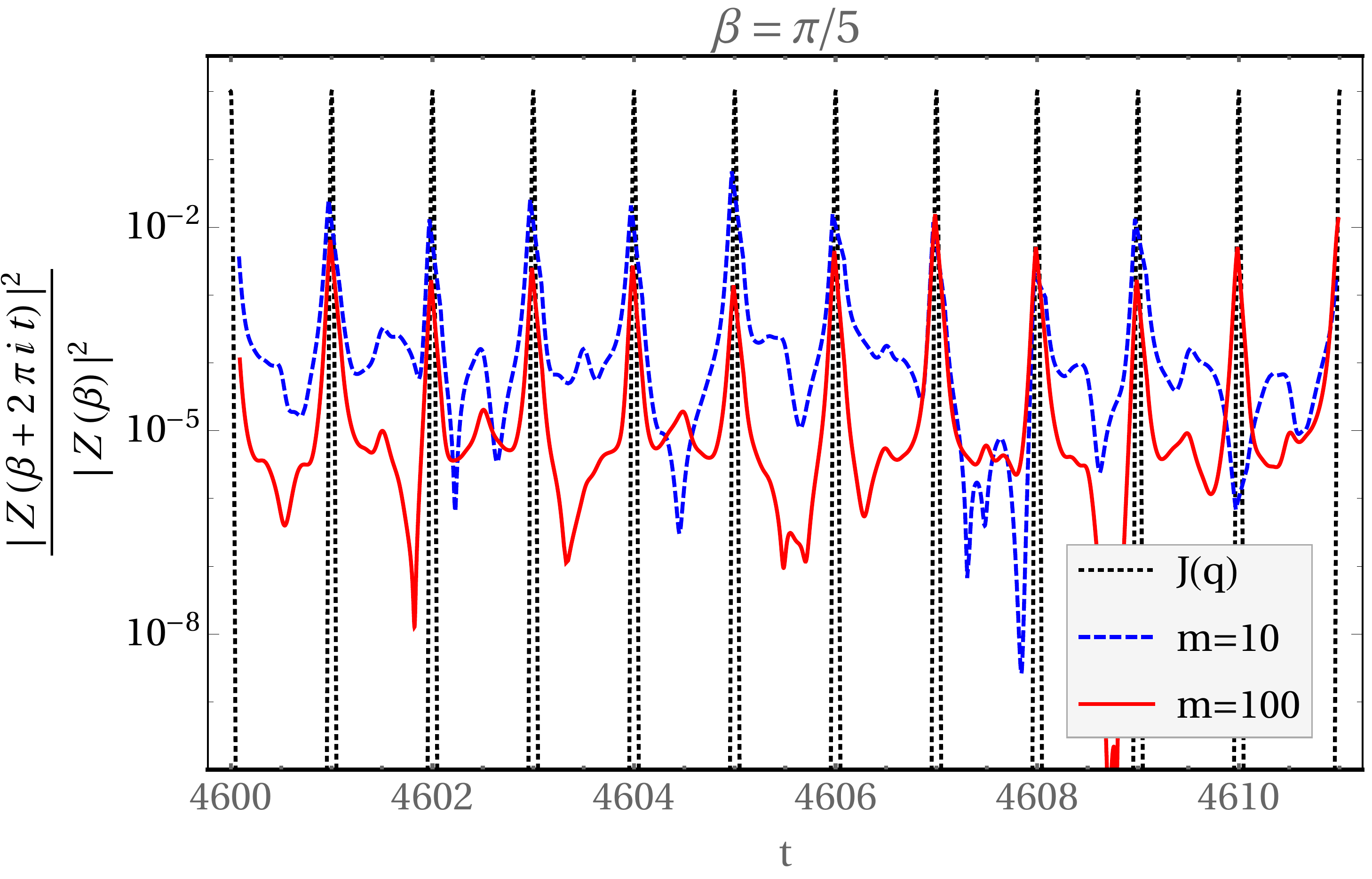}
  \caption{\label{fig:mini-recurrence}
    {\bf Left:} SFF's at continuous, early $t$, of different models.
    {\bf Right:} SFF's at continuous, late $t$, of the same models.  By eye one can see that the behavior is more irregular at late times. The dotted spikes are from the $J$-invariant, which is strictly periodic with period equal to the mini-recurrence time. The dashed and solid curves are from $m=10$ and $m=100$ minimal models respectively. The minimal model SFF's are not exactly periodic, but clearly the SFF peaks at every $t_n=2\pi n$.
  }
\end{figure}
In addition to the recurrence time, there is a ``mini-recurrence time'' $t_{\rm mrec} = 2\pi$, over which $g$ exhibits approximately periodic behavior.  This can be seen explicitly in Fig.~\ref{fig:mini-recurrence}, where we plot $g(\beta, t)$ for $\beta = \pi/5$ and $m=10$ and $m=100$.  This approximate recurrence \cite{Cardy:2014rqa, Dyer:2016pou} is a general consequence of modular invariance whenever one has an approximate notion of vacuum dominance, i.e. there is a frame where the vacuum character dominates at early times.  The reason is that  at times $t=t_n \equiv 2 \pi n$, one can always use $\sltz$-invariance of the partition function to map to $(\tau,\bar{\tau})$ from $(\tau,\bar{\tau}) =( i \frac{\beta +i t_n}{2\pi}, -i \frac{\beta +i t_n}{2\pi})$ to $(\tau,\bar{\tau}) = ( \frac{i \beta}{2\pi}, - i \frac{\beta+2 i t_n}{2\pi})$.  Under this transformation, the holomorphic part of the vacuum character is mapped to its initial time value and therefore has no late-time suppression, and the anti-holomorphic part of the vacuum character sees a modest late-time $ \propto t_n^{-3/2}$.\footnote{ By contrast, at early times  $t\ll2\pi$ the SFF  decays rapidly at high temperature,$\sim e^{\frac{8\pi^2 \b c}{12 (t^2+\b^2)}}$.} 
This contribution is therefore a large universal contribution with a mild power-law dependence, much larger than the typical values of the SFF at times in between successive $t_n$ values (where both the holomorphic and anti-holomorphic parts of characters produce large suppressions in any $\sltz$ frame).  In the case of the large-$m$ minimal models, one can also understand the mini-recurrences directly from the spectrum.  In these cases, the spectrum is known and one can check there are a large number of primaries whose conformal weights differ by integers, and therefore their respective phases align with the mini-recurrence period $t_{\rm mrec}=2\pi$.  These mini-recurrences give the dominant contribution to the partition function, so we will mostly restrict our attention to the times $t_n$.

\subsection{Dip, ramp and plateau}

\begin{figure}[t]
\begin{center}
\includegraphics[width=0.8\textwidth]{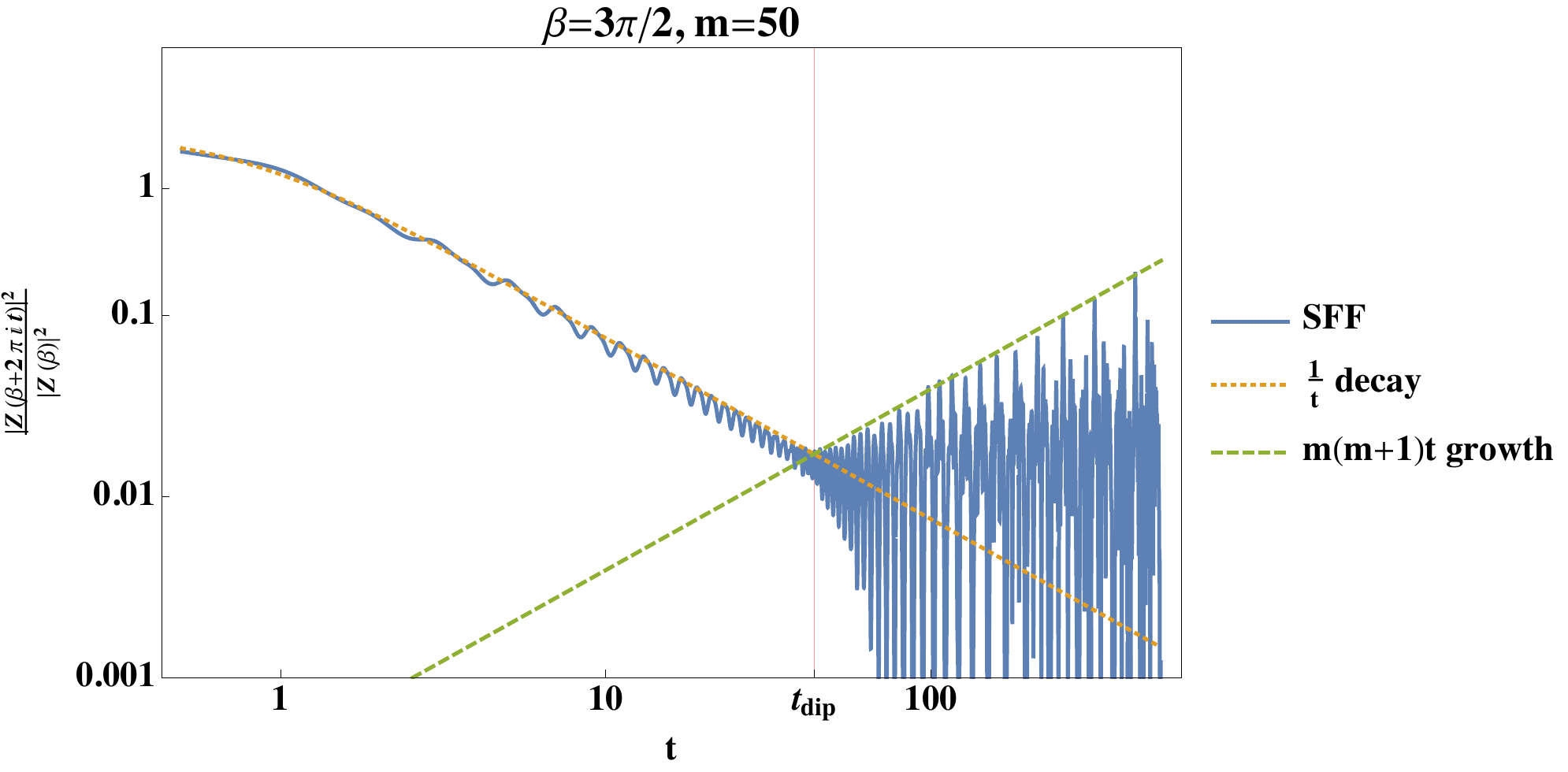}
\caption{SFF of the $m=50$ minimal model at $\beta=\frac{3\pi}2$ without any averaging. The dip and ramp behave as predicted by (\ref{eq:dipramppred}).}
\label{fig:MMSFFPlot}
\end{center}
\end{figure}

For very large $m$, the dip and ramp behavior emerge within the recurrence time.  Moreover, in this limit the behavior of the SFF is captured by the contribution of the form,
\be
g(t,\beta) = |\vartheta_3(0, e^{\frac{i\pi \tau}{m(m+1)}})|^2 .
\label{eq:minimalapprox1}
\ee
 In Appendix \ref{app:CambridgeAnalytica}, by analyzing the behavior of this $\vartheta$ function, we derive  the $1/t$ dip, linear growth ramp, and the late-time plateau in large $m$ minimal models.  We find a dip time and slope, for the normalized SFF, which scale as
\be
t_{\rm dip} \sim m \sqrt{\frac{\beta}{2\pi}}, \qquad {\rm SFF}_{\rm ramp} \sim m^2 t.
\label{eq:dipramppred}
\ee
See Figure \ref{fig:MMSFFPlot} for an explicit plot of the dip and ramp at $m=50$ and $\beta=\frac{3\pi}2$, that verifies the behavior seen in (\ref{eq:dipramppred}).

\begin{figure}[htbp]
  \centering 
  % minimal_1e4_no-avg.pdf
  % minimal_1e4_m-only.pdf
  % minimal_1e4_t-only.pdf
  % minimal_1e4_m-and-tZoomed.pdf
  \includegraphics[width=0.45\linewidth]{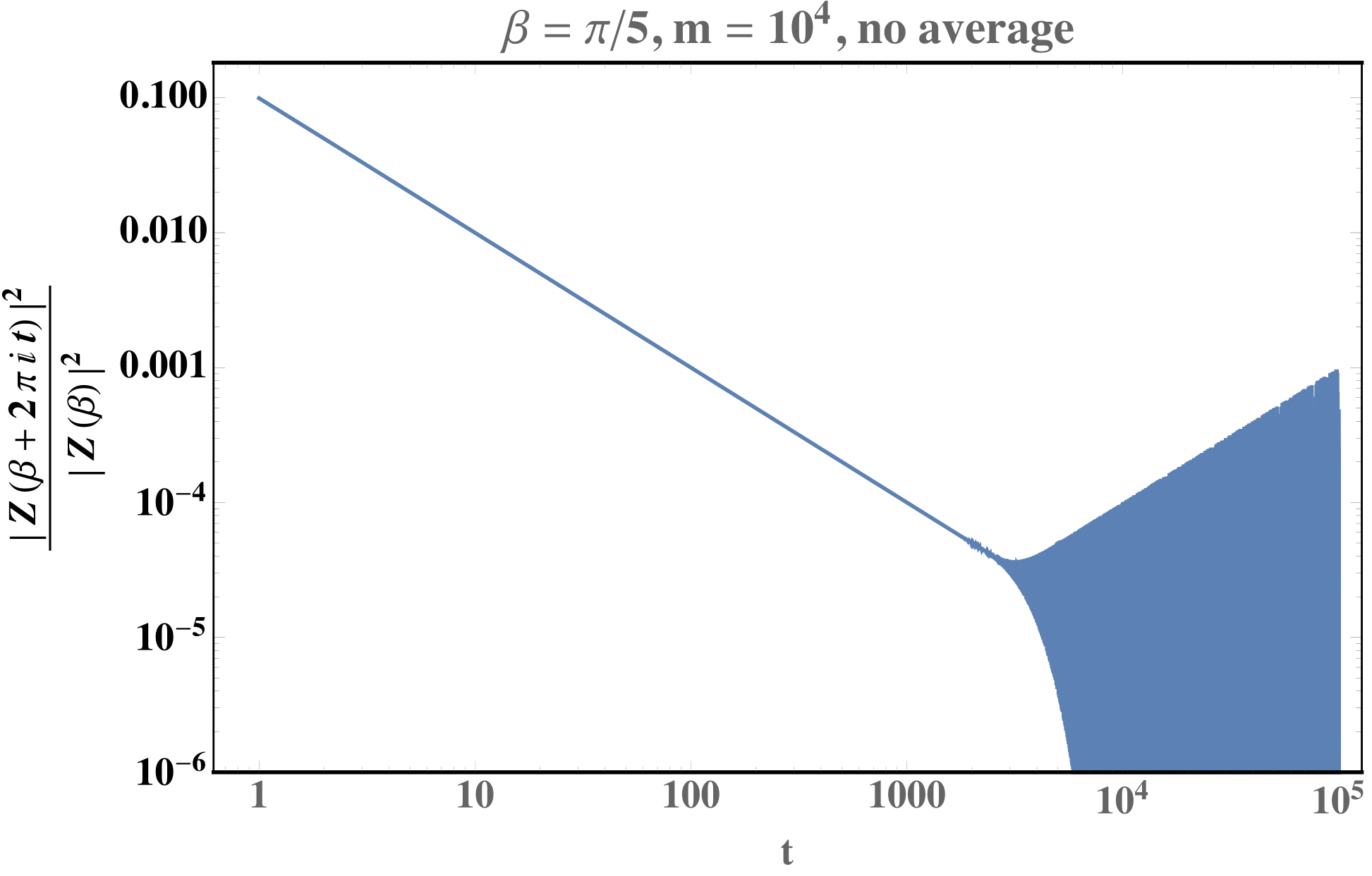}
  ~\includegraphics[width=0.45\linewidth]{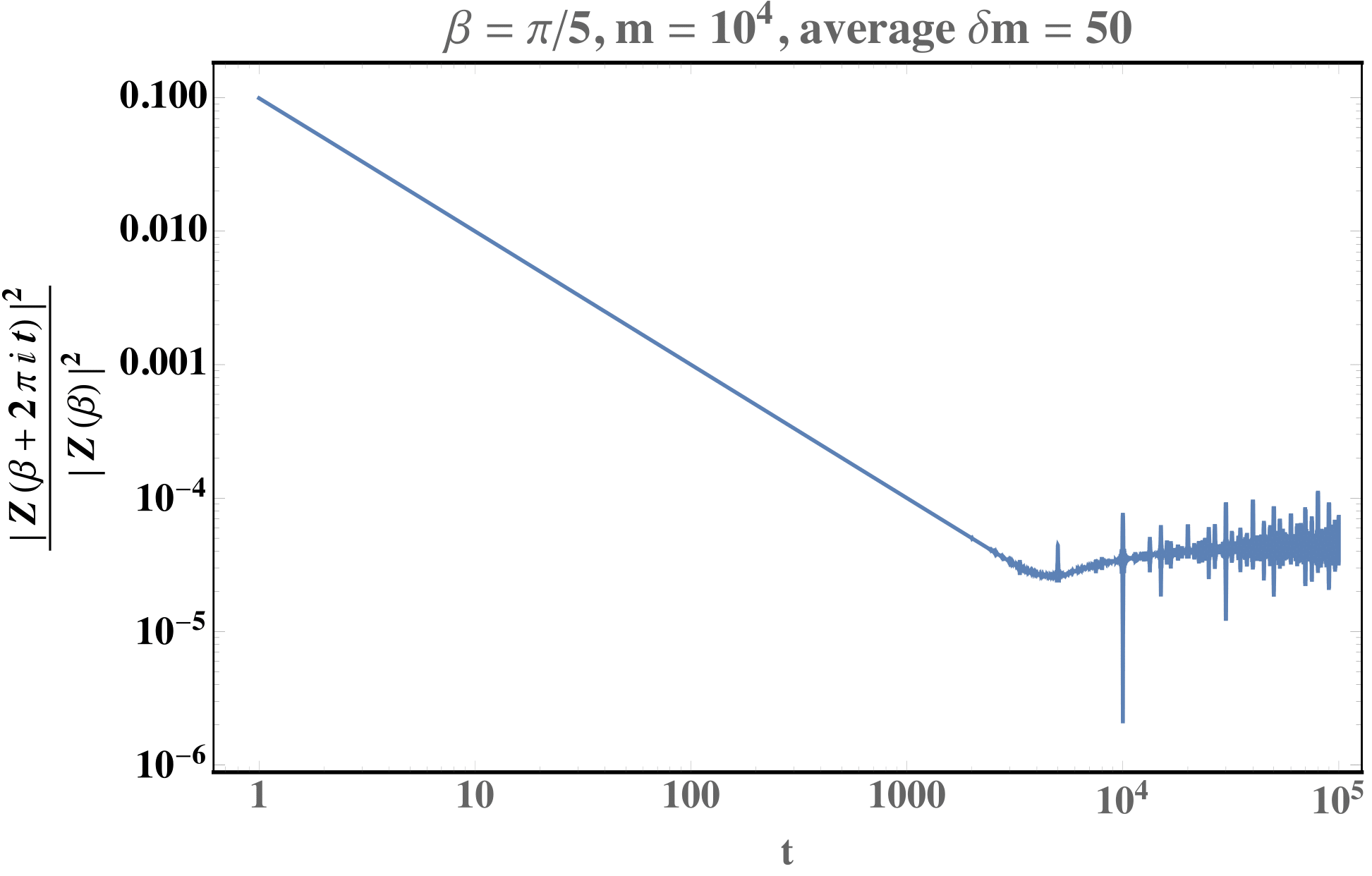} \\
  \includegraphics[width=0.45\linewidth]{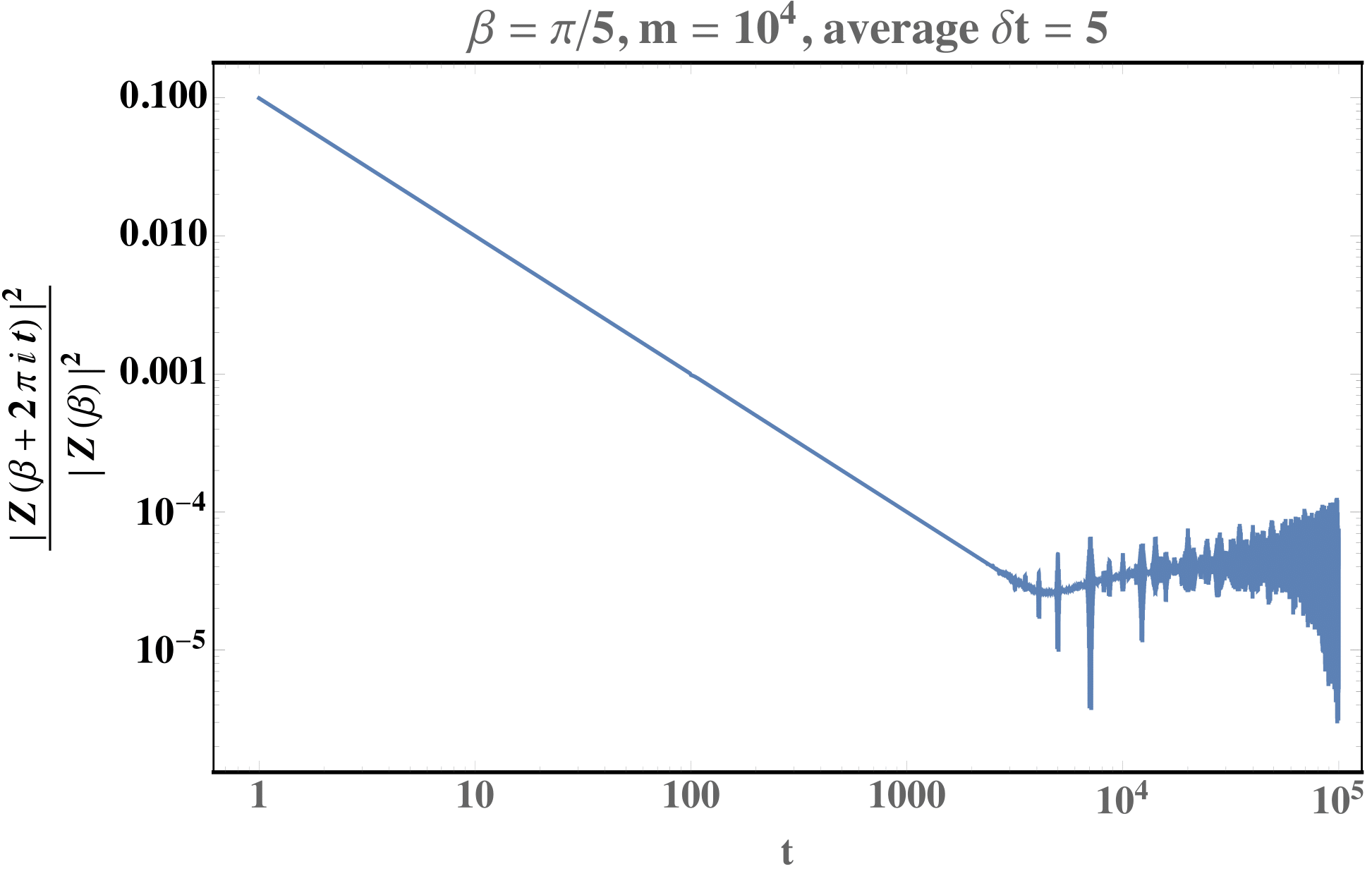}
  ~\includegraphics[width=0.45\linewidth]{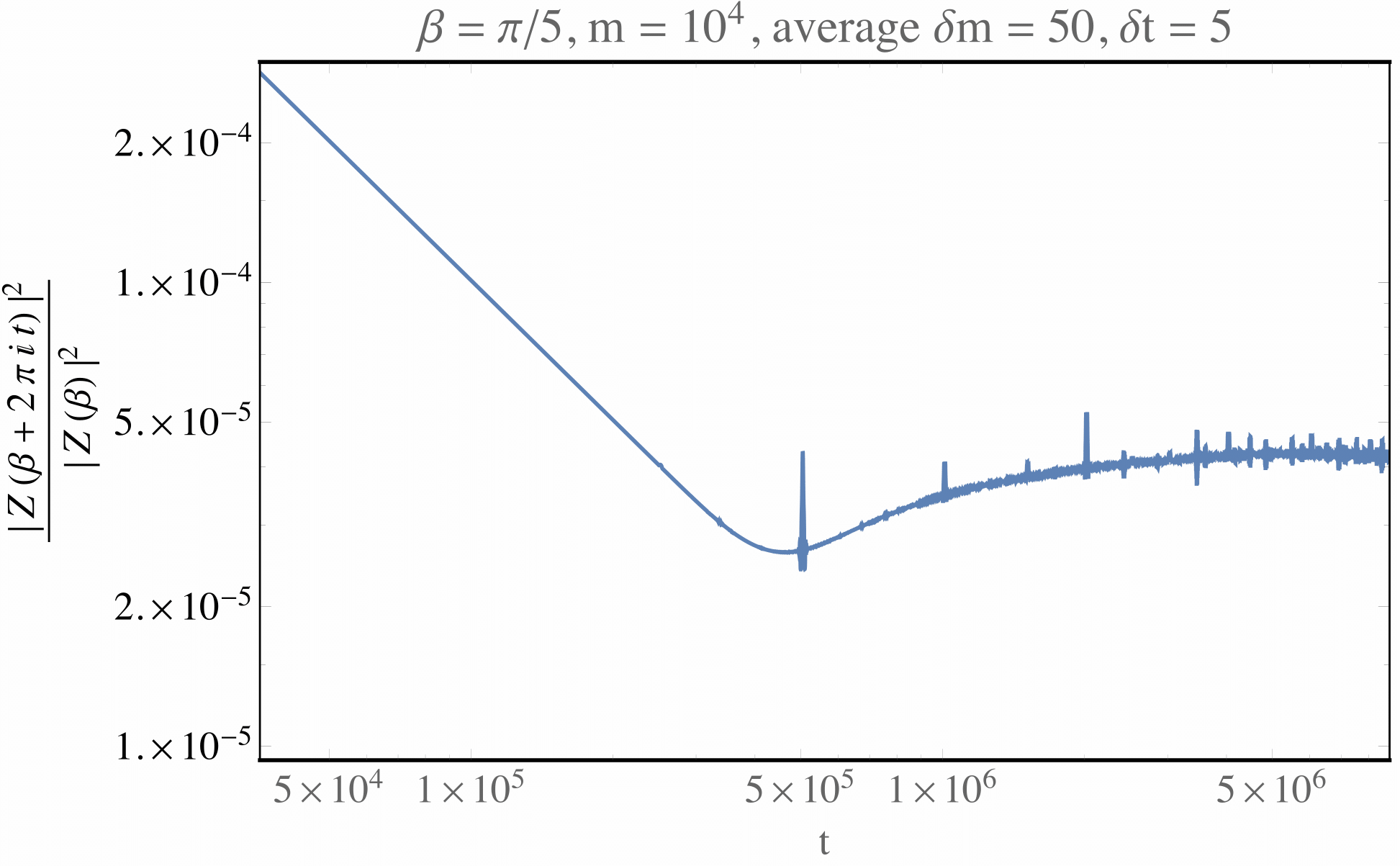}
  \caption{\label{fig:large_m_averaging}
    {\bf Upper left:} SFF of the $m=10^4$ minimal model at $\b = \pi/5$ without any averaging. The plot depicts about one-third of the full recurrence time. This unaveraged SFF has a clear dip and the late time behavior is obscured by oscillation.
    {\bf Upper right:} The same SFF only averaged over different $m$ with a window of $\pm\delta m = 50$.
    {\bf Lower left:} The same SFF only averaged over a gaussian time window of standard deviation $\delta t = 5$, for integral time greater than 100.
    {\bf Lower right:} The same SFF first averaged over different $m$, then averaged over $t$, and zoomed in on the dip and plateau.  The averaged behavior shows more clearly the ramp and plateau. 
  }
\end{figure}

We can also investigate the dip, ramp, and plateau numerically.  A single SFF of $m=10^4$, plotted in the upper-left of Fig.~\ref{fig:large_m_averaging}, shows a decay $\propto t^{-1}$. At larger times $\sim \CO(m)$, the SFF appears to be oscillatory with an overall ramp envelope.\footnote{ For random matrix theory, the smooth curve is only obtained after an average of random matrix SFF's in an ensemble. Similarly, at finite $m$, it helps to average the minimal model in a suitable ``ensemble'', which we choose to be minimal models with different $m$ within a window centered at $m_0$. As pointed out by \cite{Dyer:2016pou, Cotler:2016fpe}, the ensemble average can be replaced averaging over a parametrically small time window. Practically we find averaging both $t$ and $m$ together gives a smooth curve in the minimal model case. In Fig.~\ref{fig:large_m_averaging} we present several averaging schemes for the $m=10^4$ minimal model.}

Although minimal models are integrable, at large $m$ their SFFs have a dip-ramp-plateau structure seen in chaotic theories. Indeed the small $m$ minimal model SFFs do not exhibit this feature because of their short recurrence times. One possibility is as $m$ gets large, the SFFs ultimately are periodic, but the large number of states scramble for long enough time within a period and become approximately ergodic.

\subsection{Level statistics}

\begin{figure}[htbp]
  \centering
  \includegraphics[width=0.45\linewidth]{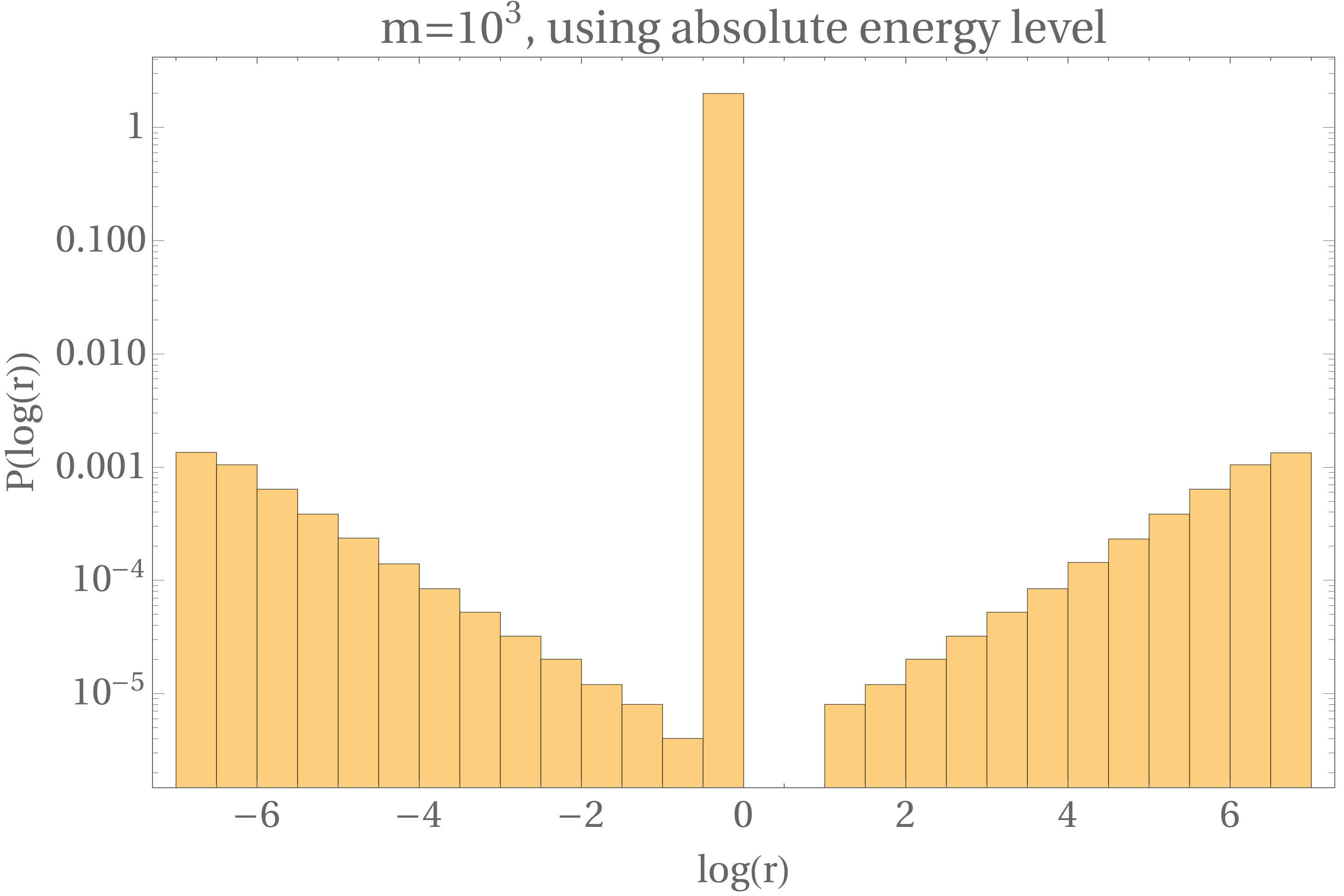} \\
  \includegraphics[width=0.45\linewidth]{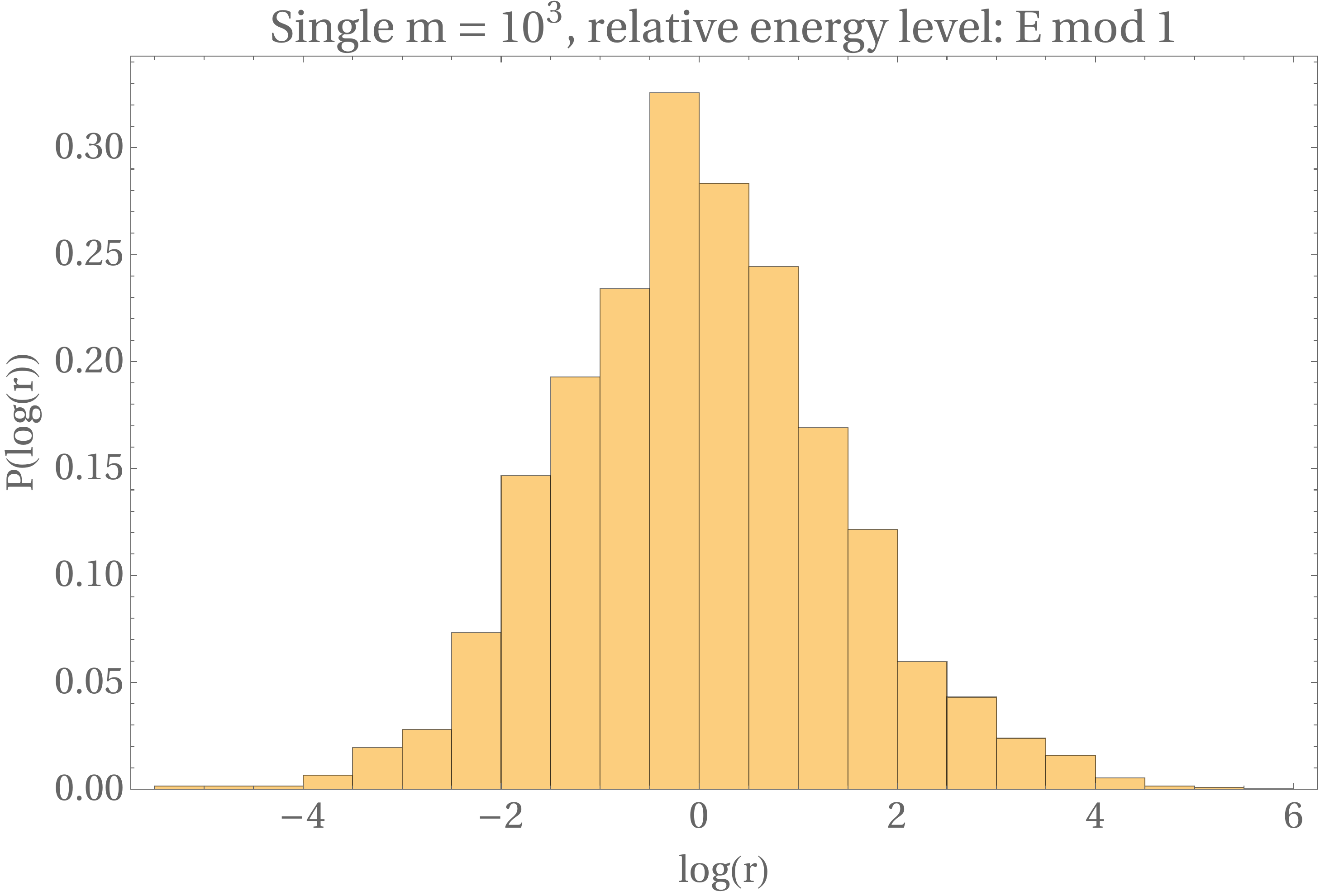}~~~
  \includegraphics[width=0.45\linewidth]{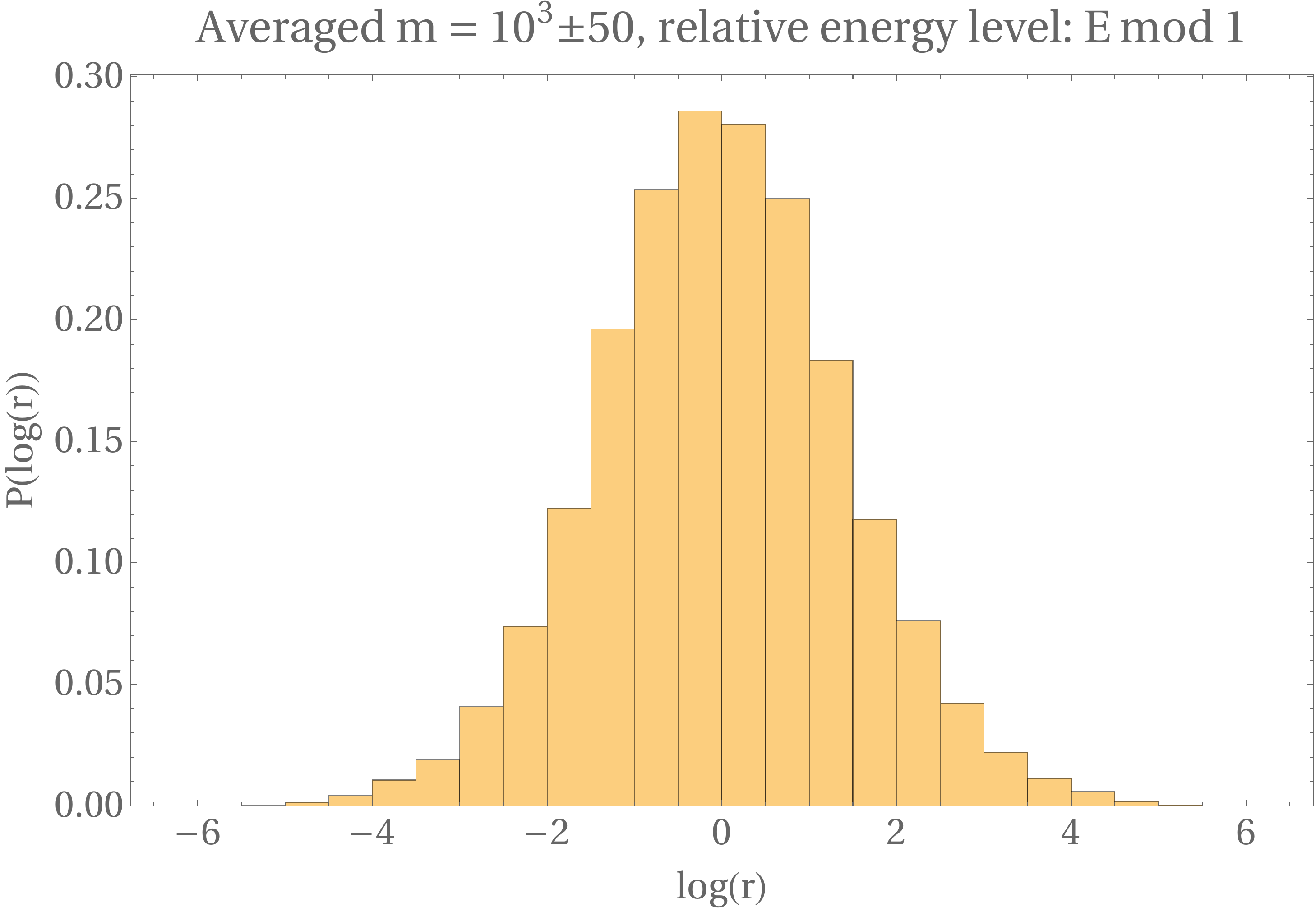}
  \caption{\label{fig:level_stat_moded}
    {\bf Upper:} Level statistics of the absolute energy levels of the $m=10^3$ minimal model.
    {\bf Lower left:} Level statistics of the $m=10^3$ minimal model with energy levels taken modulo 1 and degeneracies ignored.
    {\bf Lower right:} Same as left graph but averaging minimal models between $m=10^3-50$ to $m=10^3+50$.
  }
\end{figure}

\begin{figure}[htbp]
  \centering
  \includegraphics[width=0.45\linewidth]{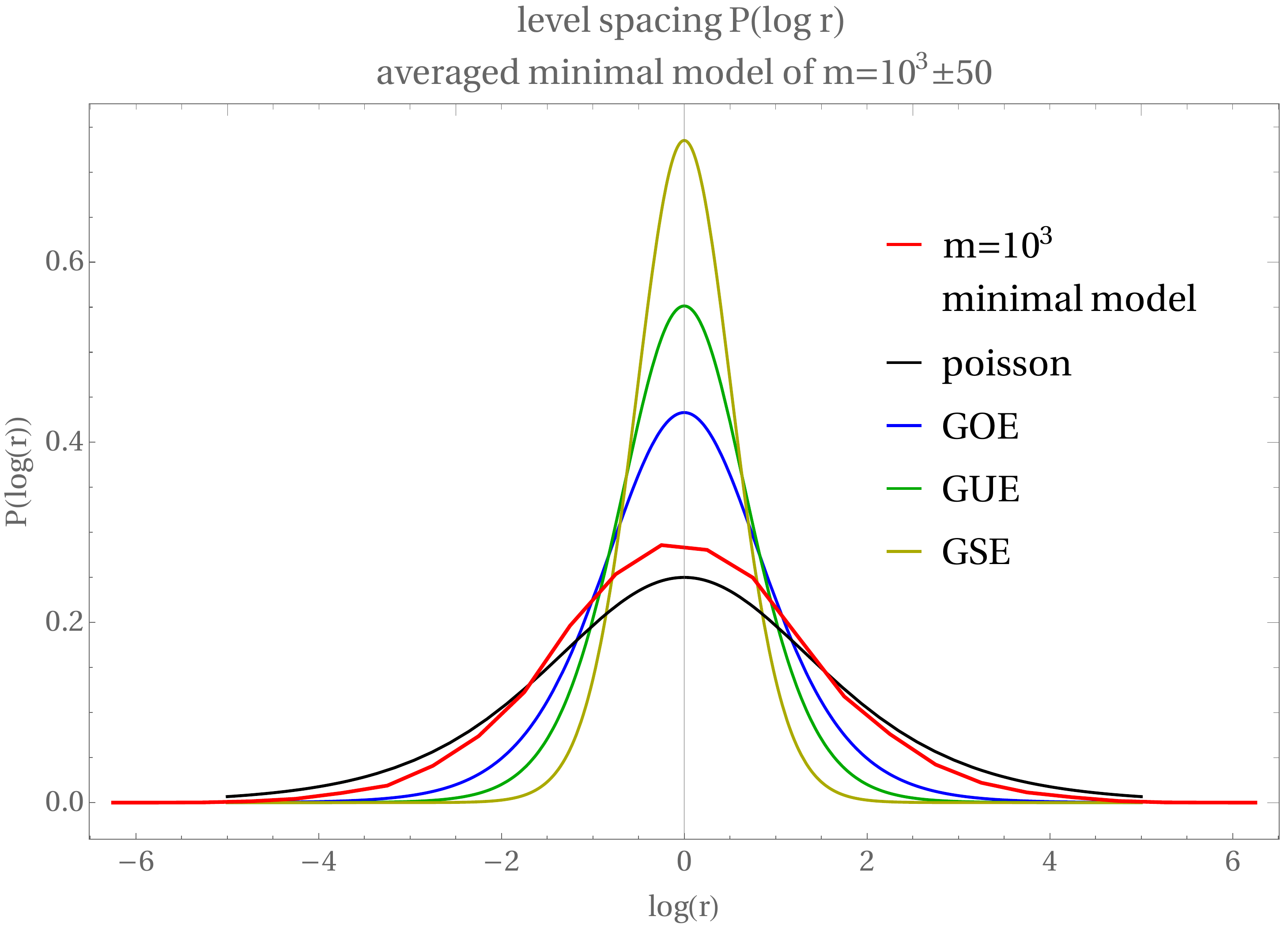}~~~
  \includegraphics[width=0.45\linewidth]{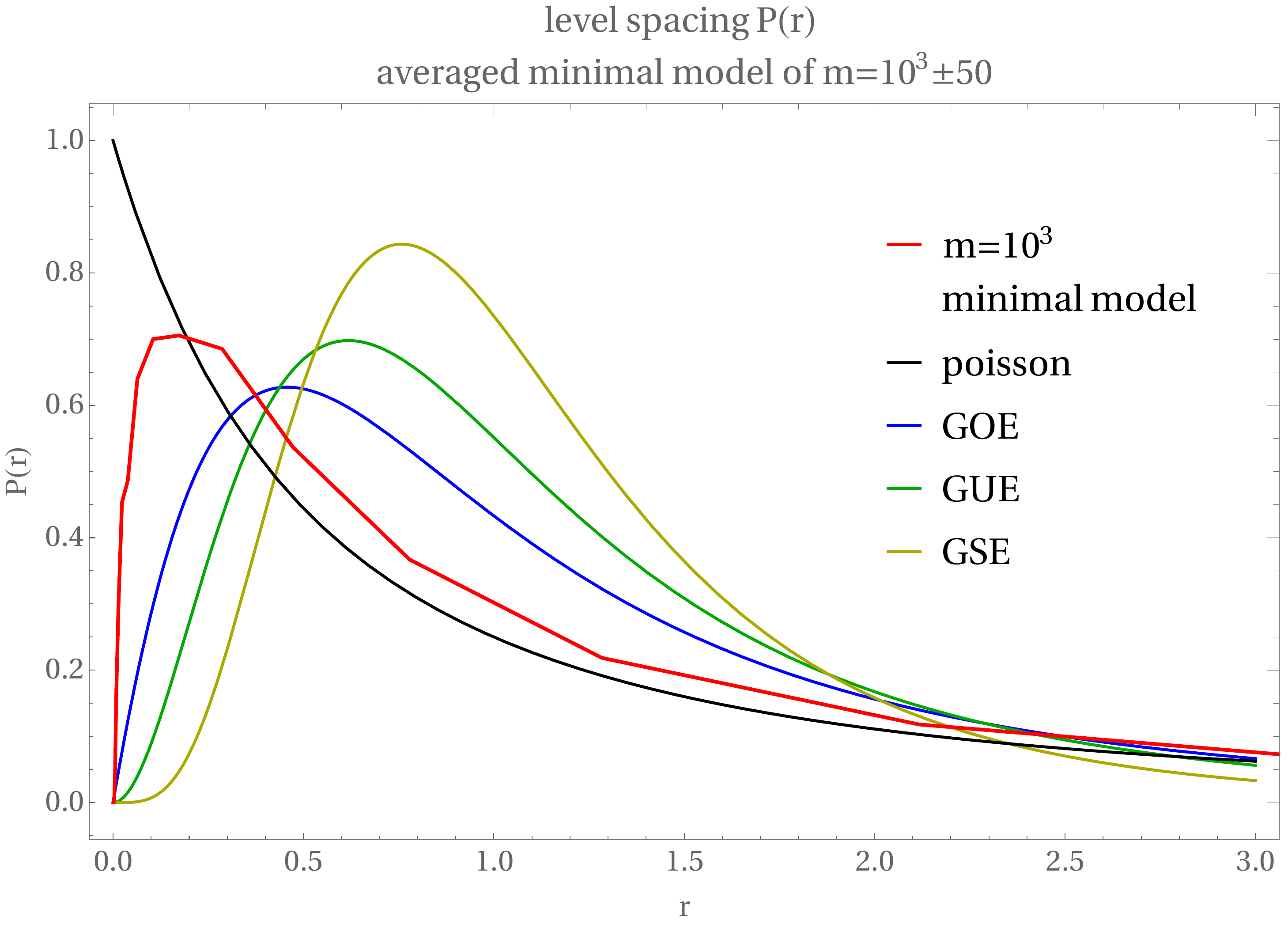}
  \caption{\label{fig:level_stat_compare}
    The smooth curve from averaging the level statistics of the $m=10^3\pm50$ minimal models with moded energy levels, compared with the Gaussian Ensembles and the Poisson distribution. A curve with $m=3 \times 10^3$ looks identical to the $m=10^3$ curve.
    {\bf Left:} Distribution as a function of log of the unfolded energy level $r$.
    {\bf Right:} Distribution as a function of the unfolded energy level $r$.
  }
\end{figure}

Since the SFF of minimal models looks somewhat similar to that of random matrix theory, this suggests a possible relationship between their spectra. One simple check is to look at their level statistics -- the distribution of nearest neighbor spacing of energy eigenvalues. 
(For a reference of level statistics see e.g. \cite{DAlessio:2016rwt}.)
Consider the energy eigenvalues in ascending order $E_n < E_{n+1}$, and define $\delta_n := E_{n+1}-E_{n}$ as the nearest neighbor differences, and $r_n := \frac{\delta_{n+1}}{\delta_n}$ as the ratio of nearest differences. The distribution of $r_n$ is independent of the overall scaling of state density and reflects the model-independent information about the state distribution.
Standard Gaussian ensembles follow the {\it Wigner Surmise} and non-chaotic models follow a Poisson distribution.

The level statistics of the $m=10^3$ minimal model are plotted in the upper graph of Fig.~\ref{fig:level_stat_moded}. The distribution does not seem to follow any of the  standard statistics. However,  2d CFTs have a special feature, namely the mini-recurrences. The SFF is dominated by $t_n = 2\pi n$. At these times, all the states $E_i$ separated by integers contribute the same phase to the SFF, $e^{i t (E_i \mod 1)}$~. We plot the level statistics of the energies mod $1$, while ignoring degeneracies, in the lower graphs of Fig.~\ref{fig:level_stat_moded}. 
Although the distribution of a single minimal model at $m=10^3$ is noisy, a smooth curve can be obtained by averaging over a parametrically small window $\delta m = 50$ around $m_0=10^3$. The smooth curve interpolates between those of the Gaussian ensembles and Poisson distribution, as shown by Fig.~\ref{fig:level_stat_compare}, suggesting that large $m$ minimal models have weak level repulsion\footnote{We ignored the degeneracy of states. A more principled approach would be to factor out some internal symmetry of the model.}.

The minimal models have both small $c$ and small gap. The bulk dual theories are not Einstein gravity. Raising the gap makes the models more ``gravity-like'', and since the minimal models already exhibit interesting chaotic behavior, one hopes the large-$m$ models with large $c$ and large gap are in better agreement with gravity. In the next subsection we introduce the mathematics that generalizes to these models.

\section{Generalizing to Large \texorpdfstring{$c$}{c} and Large Gap}
\label{sec:modular}

Every two-dimensional rational conformal field theory (RCFT) provides a natural set of vector-valued modular forms: the characters of the primary operators. In particular, these are holomorphic functions on the upper half plane that transform under a finite-dimensional representation of $\sltz$. If we call these characters $\chi^\mu(\tau)$, where $\mu$ runs over the representation index of $\sltz$, the partition function of the RCFT is given by
\be
Z(\tau, \bar\tau) = \mathcal{M}_{\mu \nu} \chi^{\mu}(\tau) \chi^{\nu}(\bar\tau)
\label{eq:pfrcftgen}
\ee
where $\mathcal{M}$ commutes with both the representation of $S$ (which takes $\tau\rightarrow-\frac1\tau$) and the representation of $T$ (which takes $\tau\rightarrow\tau+1$). For simplicity we will take $\mathcal{M}$ to be the identity matrix, but other possibilities may exist. 

Unitarity provides further constraints on the behavior of the characters $\chi^\mu(\tau)$. If the central charge of the RCFT is $c$, then the vacuum character  goes as $q^{-\frac{c}{24}} + \ldots$, and this is the highest order pole in any component of the $\chi$ vector. Moreover, the $q$-expansion of $\chi^{\mu}(\tau)$ must give non-negative integers for all components of $\chi$. In this section, we describe an algorithm borrowed from \cite{Bantay:2005vk} that does the following. Given a finite-dimensional representation of $\sltz$, a set of $\chi^\mu(\tau)$ is constructed that transforms as a vector-valued modular form, and satisfies the unitarity constraints necessary to be a CFT partition function at arbitrarily large central charge. In particular, the central charge $c$ can be taken large independently of the $\sltz$ representation dimension. Some technical details of the construction are relegated to Appendix \ref{sec:bgapp}.

\subsection{Bantay-Gannon vector-valued modular forms}
\label{sec:vModForm}

In this section we will simply state the results of \cite{Bantay:2005vk} (see also \cite{Bantay:2007zz, Bantay:2011}). Consider any $r$-dimensional representation of $\sltz$, which is generated by two elements: $S$ and $T$. We choose a basis where $T$ is diagonal and is given by $T = \exp \left( 2 \pi i \Lambda \right),$ where $\Lambda = \text{diag}(\lambda_1,\,\lambda_2,\,\cdots, \lambda_r).$ Then for any positive integers $i, j$ with $1 \leq i \leq r$, can find a vector-valued function of $\tau$ in the upper half plane, $\X^{(i; j)}$, such that
\begin{equation}
\X^{(i; j)} = q^{\Lambda} \begin{pmatrix} \mathcal{O}(q^0) \\ \ldots \\ \mathcal{O}(q^0) \\ q^{-j} + \mathcal{O}(q^0) \\ \mathcal{O}(q^0) \\ \ldots \\ \mathcal{O}(q^0)\end{pmatrix}
\end{equation}
where the $q^{-j}$ pole occurs in the $i^{\text{th}}$ slot of the vector.

\subsection{Unitarity and holomorphicity}
\label{sec:unitary-holo}
From the technology developed in \cite{Bantay:2005vk}, we can explicitly construct modular-invariant and positive-definite partition functions that (a) transform under arbitrarily large representations of $\sltz$, (b) correspond to potential CFTs with arbitrarily large central charge $c$, and (c) have a gap to the first nontrivial primary at $\mathcal{O}(c)$ above the vacuum.

In particular, for any representation and any nonnegative integer $k$, consider the vector given by
\be
\chi = \sum_{i=0}^{k} d_i \X^{(1; k-i+1)}
\ee
where $d_i$ satisfies
\be
\sum_{i=0}^{\infty} d_i q^i = \prod_{n=2}^{\infty} \frac1{1-q^n}.
\ee
By construction, the function
\be
Z(\tau, \bar\tau) = \chi^\mu(\tau) \chi_\mu(\bar\tau)
\label{eq:zttb}
\ee
is a potential partition function for a CFT with central charge $c=24(k+1-\lambda_1)$. In the large $k$ limit, the first nontrivial primary occurs $\sim \frac c{24}$ above the vacuum.

To have the partition function $Z(\tau,\bar\tau)$ transform in arbitrarily large representations of $\sltz$, we simply need to choose a family of RCFTs with arbitrarily large number of primary operators. For simplicity we have chosen to focus on the unitary Virasoro minimal models, whose characters provide a family of $\frac{m(m-1)}2$-dimensional representations of $\sltz$.

The only remaining question is if the $q, \bar q$-expansion of $Z(\tau, \bar\tau)$ in (\ref{eq:zttb}) gives manifestly positive terms. We can estimate the coefficient in front of the $q^{\Delta}$ term in slot $i$ of $\chi^\mu$ by using a vector-valued Rademacher expansion. (See, for instance, (2.17) of \cite{Dijkgraaf:2000fq}.) At large $k$, for $\Delta>0$, the sign of the coefficient in front of $q^{\Delta}$ is solely determined by the sign of $S$-matrix component $S_{1,i}$. From an explicit form of the modular transformation properties of the minimal model characters, it can be shown that this is always positive.\footnote{Note that for the constant pieces, that is $\Delta=0$, the first term of the Rademacher sum may not necessarily dominate, in which case our argument breaks down. If the $\Delta=0$ term happens to be negative, we can always add a few polar terms very close to threshold to push it to be positive. In the examples that we have checked, this procedure always works; it would be interesting to find an analytic proof of this statement.}

\subsection{A larger gap in \texorpdfstring{$c$}{c}?}
\label{sec:c12maybe}

We end this section with a potentially interesting question. In Section \ref{sec:unitary-holo}, we constructed an explicit family of positive-definite partition functions with a gap of $\frac{c}{24}$ that transform in arbitrarily large representations. It would be interesting if we could get a gap larger than $\frac{c}{24}$ -- the BTZ black hole mass suggests that a gap of $\frac{c}{12}$ may be possible, for instance.

If we ignore positivity, we will see that we can get an arbitrarily large gap in $c$. This is in contrast to the holomorphic case, where the space of modular functions is constraining enough so that even without positivity, one cannot do better than a gap of $\frac{c}{24}$.

Consider a partition function coming from a candidate CFT transforming under a $r$-dimensional representation of $\sltz$ with central charge $c = 24k + \mathcal{O}(1)$. We would like to see, given $r$ and $k$, how large we can make the gap $\Delta$ to the first nontrivial primary. In other words, we are trying to build a partition function that satisfies
\be
Z(\epsilon q, \epsilon \bar q) = \epsilon^{-\frac{c}{12}} q^{-\frac{c}{24}}\bar{q}^{-\frac{c}{24}} \(\prod_{n=2}^{\infty} \frac{1}{(1-\epsilon^nq^n)(1-\epsilon^n\bar q^n)} + \mathcal{O}(\epsilon^{\Delta})\).
\label{eq:thedream}
\ee

First let us count the number of degrees of freedom we have to utilize. We are allowed to tune the numbers multiplying any of the vectors $\X^{(\xi;n)}$ for $1\leq\xi\leq r$ and $1\leq n \leq k$, for both the holomorphic and anti-holomorphic piece of the partition function giving us $r^2k^2$ degrees of freedom.

Now let us look at the number of terms we have to tune to match (\ref{eq:thedream}).  Recall in our $r$-dimensional representation of $\sltz$, we have
\be
T = \exp \left( 2 \pi i \Lambda \right),~~~\text{where}~~ \Lambda = \text{diag}(\lambda_1,\,\lambda_2,\,\cdots, \lambda_r)~.
\ee
Suppose there are $r'$ distinct $\lambda_i$'s mod 1. In each of the $r'$ ``sectors" of the partition function, we have a total of $\sim\frac{\Delta^2}2$ terms to match (coming from the $\epsilon^{0-\frac{c}{12}}, \epsilon^{1-\frac{c}{12}}, \ldots, \epsilon^{\Delta-\frac{c}{12}}$ terms; the $\epsilon^{i-\frac{c}{12}}$ term has $i$ possibilities distributed between $q$ and $\bar q$). Thus we have $\frac{r' \Delta^2}2$ terms to match. If we set these equal we get
\be
r^2 k^2 = \frac{r' \Delta^2}2.
\ee
Since $r \geq r'$ by definition, we can thus get a gap (ignoring positivity) of
\be
\Delta \geq \sqrt{2r}k + \mathcal{O}(1) = \frac{\sqrt{2r}}{24}c + \mathcal{O}(1)
\ee
at large $c$. By making $r$ extremely large, we can get an arbitrarily large gap in $c$. Of course, we know from e.g. \cite{Hellerman:2009bu} that this function cannot be positive-definite. This result is reminiscent of \cite{Maloney:2007ud}, where a naive $\sltz$ sum of the vacuum character gave terms with negative degeneracy. It would be interesting to see if for certain representations, we can get a family of positive-definite functions at large $c$ that has a gap larger than $\frac{c}{24}$ (see \cite{Maloney:2007ud, Keller:2014xba} for a construction of a partition function with a gap of $\frac{c}{12}$, although with a continuous spectrum of states).

\section{Analysis of the Large Gap SFFs}
\label{sec:largegapsff}

The spectral form factor (SFF), related to an analytic continuation of the partition function and defined in \eqref{sffdef}, is interestingly sensitive to the spectral structure of a theory and can serve as a useful probe of integrable versus chaotic behavior. 
In this section we first review properties of the SFF and then use it to examine our construction of large $c$ large gap theories, see  e.g. \cite{Cotler:2016fpe} and references therein for more details.

A prototypical non-integrable spectral form factor  is characterized  by three different phases, dubbed the \textit{dip}, \textit{ramp}, and \textit{plateau} in \cite{Cotler:2016fpe}. This is the structure of the SFF in random matrix theory and in the Sachdev-Ye-Kitaev model \cite{sachdev1993gapless,Kitaev:2017awl,Cotler:2016fpe}. This basic form has also been conjectured to be present generically in non-integrable 2d CFTs with large gaps \cite{Dyer:2016pou}. These three time periods of the SFF probe different aspects of the spectrum.
\begin{itemize}
\item{\textbf{Plateau}}: The plateau sets in at times $t \gg e^{S(2\beta)}\sim e^{\frac{8\pi^{2}}{\beta}\frac{c}{24}}$ and is present in any theory with a discrete spectrum. Its average height can be estimated as
\es{averagelatetimeheight}{
\lim_{t\rightarrow\infty}\frac{1}{t}\int_{0}^{t}g(\beta,t^{\prime})dt^{\prime}&= \sum_{i}N^{2}e^{-2E_{i}\beta}
\geq Z(2\beta)\sim e^{\frac{4\pi^{2}}{\beta}\frac{c}{24}}.
}

\item{\textbf{Ramp}}: The growth of the SFF approaching the plateau is called the ramp, and probes the correlation structure of closely separated energy levels. In RMT the pair correlation is given by the sine kernel, giving a linear growth to the SFF. This linear growth of the SFF is generic in chaotic theories and has relatively recently been appreciated in theories dual to gravity.

\item{\textbf{Dip}}: The dip probes correlations between widely separated eigenvalues and there is no known universality of this behavior. In order for the dip-ramp-plateau structure to be visible, the SFF must fall off as at least a power law in $t$. 
\end{itemize}

In \eqref{averagelatetimeheight} we characterized the late time behavior by a continuous average, it is convenient to also compute the SFF averaged over integer times.
\es{intavgg}{
\bar{g}(\beta,n)&\equiv\frac{1}{n}\sum_{n^{\prime}=0}^{n}g(\beta, t_{n^{\prime}}) .
}

\begin{figure}
  \centering
  \includegraphics[width=0.8\linewidth]{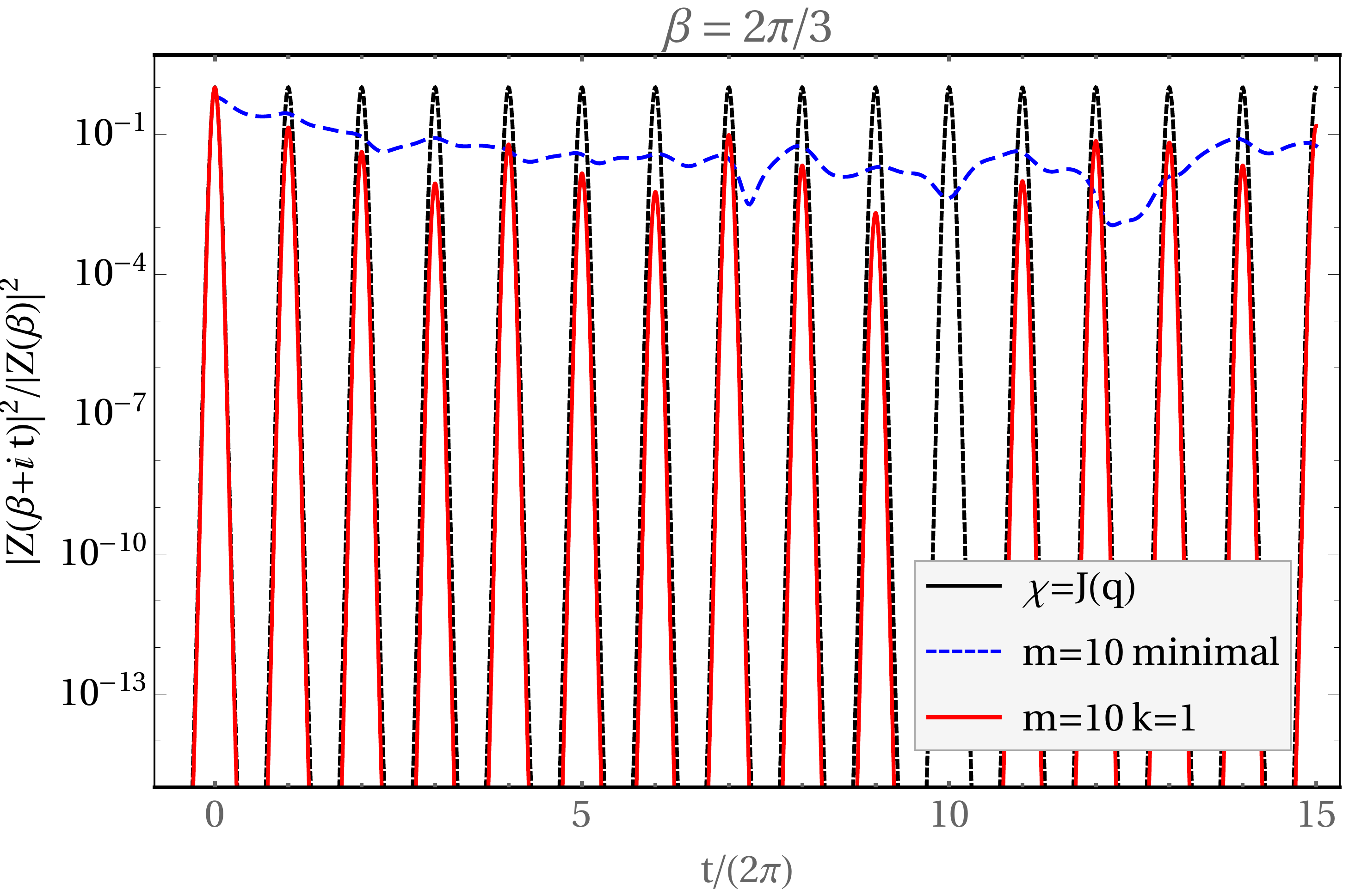}
  \caption{
    \label{fig:m_10_gappy_exact_cont}
    A plot of the spectral form factor of the $m=10$, $k=1$ model at continuous time $t$ (red solid), compared with the $J$-function (black short dashed) and the $m=10$ minimal model (blue long dashed). The function is clearly dominated by peaks at integer times.
  }
\end{figure}

This average is nice for two related reasons. First, because of the mini-recurrences, the dominant contributions to the averaged SFF come from integer times $t=t_n \equiv 2\pi n$. See Fig. \ref{fig:m_10_gappy_exact_cont} for an example. Second, modularity helps simplify $\bar{g}$ as integer time evolution is related to a modular transformation: integer time evolution is equivalent to a separate $\sltz$ transformation on $\tau$ and $\bar{\tau}$.

In the particular constructions we are considering in the paper, our partition functions are built out of characters that transform under a finite dimensional representation of $\sltz$, and thus time evolution at integer times can be mapped to matrix evolution under a finite dimensional matrix.\footnote{Note that we are abusing notation somewhat and have changed the argument of $\chi$ from $\tau$ to $\beta$, since this will make the formulas more transparent and hopefully it will be clear from context which is meant.}

\es{matrixevolution}{
Z(\beta,t_{n})&=\chi^{\mu}\left(\frac{4\pi^{2}}{\beta+it_{n}}\right)\chi_{\mu}\left(\frac{4\pi^{2}}{\beta+it_{n}}\right)\\
&=S_{\mu\alpha}(T^{2n})^{\alpha\beta}S_{\beta\nu}\chi^{\mu}\left(\frac{4\pi^{2}}{\beta}\right)\chi^{\nu}\left(\frac{4\pi^{2}}{\beta}\right)
} 

 Below we will use this matrix evolution of the partition function and the integer averaged SFF, $\bar{g}$, to discuss two different phases of large $c$, large gap theories.
The constructions we have described so far have a few tunable parameters.
\begin{itemize}
\item $c$ - the central charge.
\item $\Delta$ - the gap in the partition function (in this section we will take this to be $\sim c/24$).
\item $m$ - the minimal model index, i.e. the diagonal minimal model $\mathcal{M}(m, m+1)$.\footnote{In reality the discussion in this section, and much of the paper, only relies on having a family of finite dimensional representations with growing dimension and degree of irrationality, not on the details of the particular representations associated to the minimal models. The minimal model index $m$ is really a proxy for the dimension of the representation $d=\frac{m(m-1)}{2}$.}
\end{itemize}
We will look at both the gravity-like regime, $m\gg c$, and the more numerically tractable $m < c$.

\subsection{Gravity-like regime}
We are interested in making contact with large $c$ theories that capture as much of the spectral structure of Einstein gravity as possible. In such theories, the recurrence time should be much larger then any of the dip-ramp-plateau timescales we have discussed above. In our constructions, the recurrence time scales parametrically in $m$ and so we should have
\es{rectimescale}{
c \sim \Delta \gg  1\\
m^{2} \gg e^{c}
} 

It is quite difficult to construct the conformal vectors or partition functions explicitly, or even numerically for this range of parameters, due to the large representation dimension. We can make a bit of progress studying the properties of the SFF analytically. It is convenient to consider separately the matrix evolution corresponding to the vacuum and heavy states in the representation, and to normalize by the vacuum contribution at $t=0$. 

\es{SFFsplit}{
\mathbf{\hat{Z}}&=\underbrace{S_{0\alpha}T_{\alpha\beta}^{2n}S_{\beta 0}}_{V\times V}+\underbrace{2S_{0\alpha}T_{\alpha\beta}^{2n}S_{\beta i}\hat{\chi}^{i}}_{V\times H} + \underbrace{S_{i\alpha}T_{\alpha\beta}^{2n}S_{\beta j}\hat{\chi}^{i}\hat{\chi}^{j}}_{H\times H}
}

Here, $\hat{\chi}^{i}$ is the $i^{\text{th}}$ (heavy) character normalized by the vacuum character: $\hat{\chi}^{i}=\chi^{i}/\chi^{0}$.

We can say a few things analytically about each of these pieces.

\begin{itemize}
\item{$\mathbf{V\times V}$:} At $t=0$ this piece of the partition function exponentially dominates. This is the usual statement that at high temperatures the modular $S$ image of the vacuum dominates. We have normalized things so that this contribution is equal to 1 initially and all other pieces start at $e^{-\frac{\# \Delta c}{\beta}}$. 

As time grows, this contribution decays. At very late times we have, averaging over integer times,
\es{vvatlate}{
\lim_{n\rightarrow\infty}\frac{1}{n}\left|\sum_{n^{\prime}=0}^{n}V\times V\right|^{2}&=\lim_{n\rightarrow\infty}\frac{1}{n}\left|\sum_{n^{\prime}=0}^{n}S_{0\alpha}T_{\alpha\beta}^{2n^{\prime}}S_{\beta 0}\right|^{2}\\
&\sim\sum_{\alpha}S_{0\alpha}^{4} \,\propto \, \frac{1}{m^{2}}\,.
}
The last relation is computed explicitly in Appendix \ref{app:vv}. With our scaling of $m$ this contribution is exponentially suppressed relative to its initial value at large $c$ and also much smaller then the final plateau.

This $m^{2}$ suppression is not restricted to late times, but shows up almost immediately; indeed we can compute the $V\times V$ contribution to the partition function at finite times (see \eqref{vvearly}) and we also have:
\es{vvatearly}{
\left|V\times V\right|^{2}\propto\frac{1}{m^{2}n}\,.
}
\item{$\mathbf{H \times H}$:}
As a result of the large gap, the $H\times H$ contribution is small for all time. To see this, first note that the $\hat{\chi}_{i}$ are each down by a factor of $e^{-\frac{4\pi^{2}}{\beta}\Delta}$. One may worry, that the sum over the $m^{2}-1$ components of $\hat{\chi}_{i}$  give an extra enhancement, but this sum converges and decreases with decreasing $\beta$. We can thus bound the heavy heavy contribution at high temperatures (see \eqref{hhcont} for details)
\es{hhuniformbound}{
\lim_{\beta\rightarrow 0}\left|H\times H\right| \leq C e^{-\frac{8\pi^{2}}{\beta}\Delta}\,.
}
for some time independent number, $C$. In this normalization, $\bar g_{\text{plateau}} \sim  e^{-\frac{4\pi^{2}}{\beta}\Delta}$, so the $H\times H$ piece is small for all time, and cannot be the dominant contributor to the plateau.

\begin{figure}[t!]
\begin{center}
\includegraphics[width=0.45\textwidth]{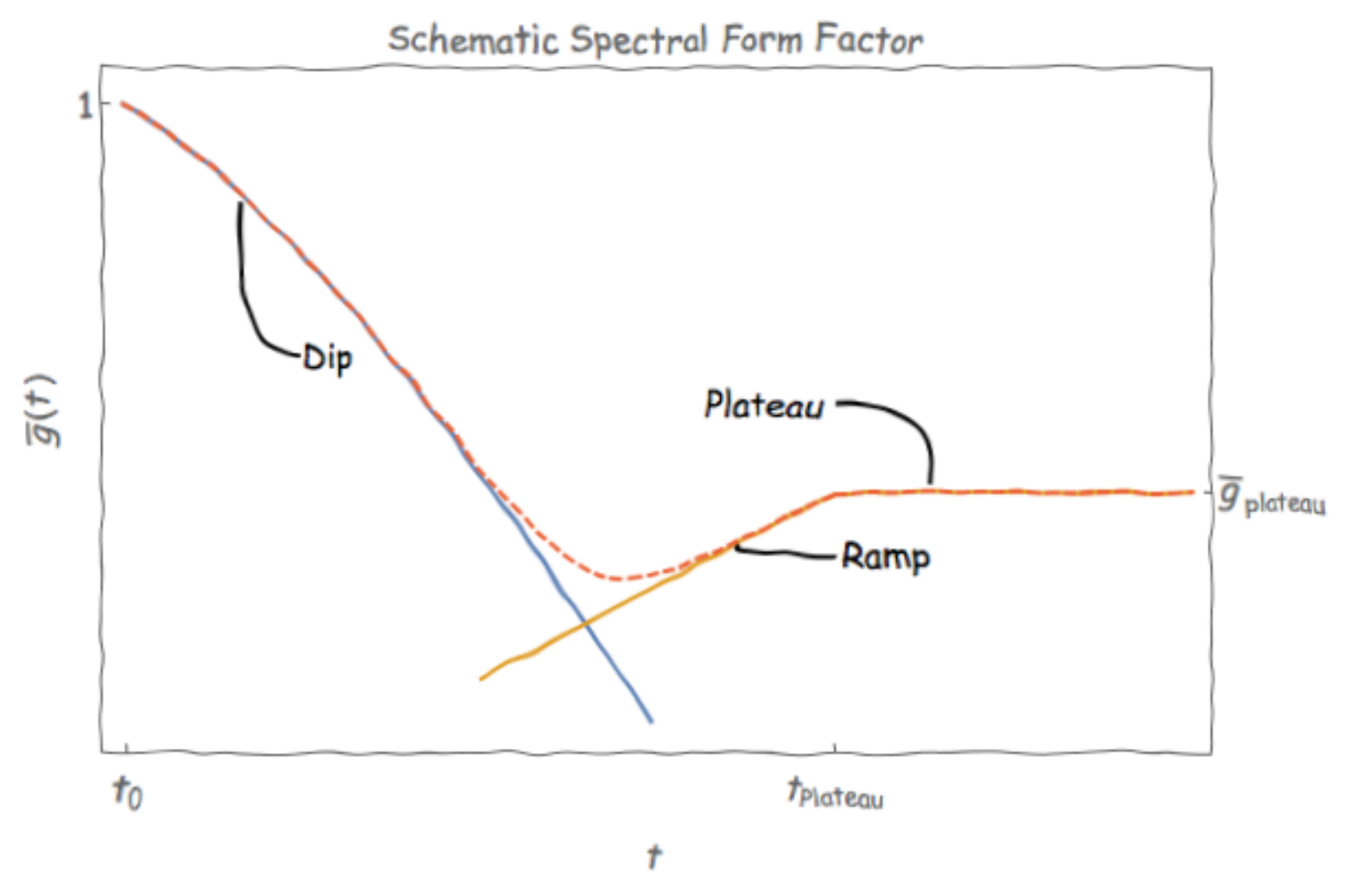}
\includegraphics[width=0.45\textwidth]{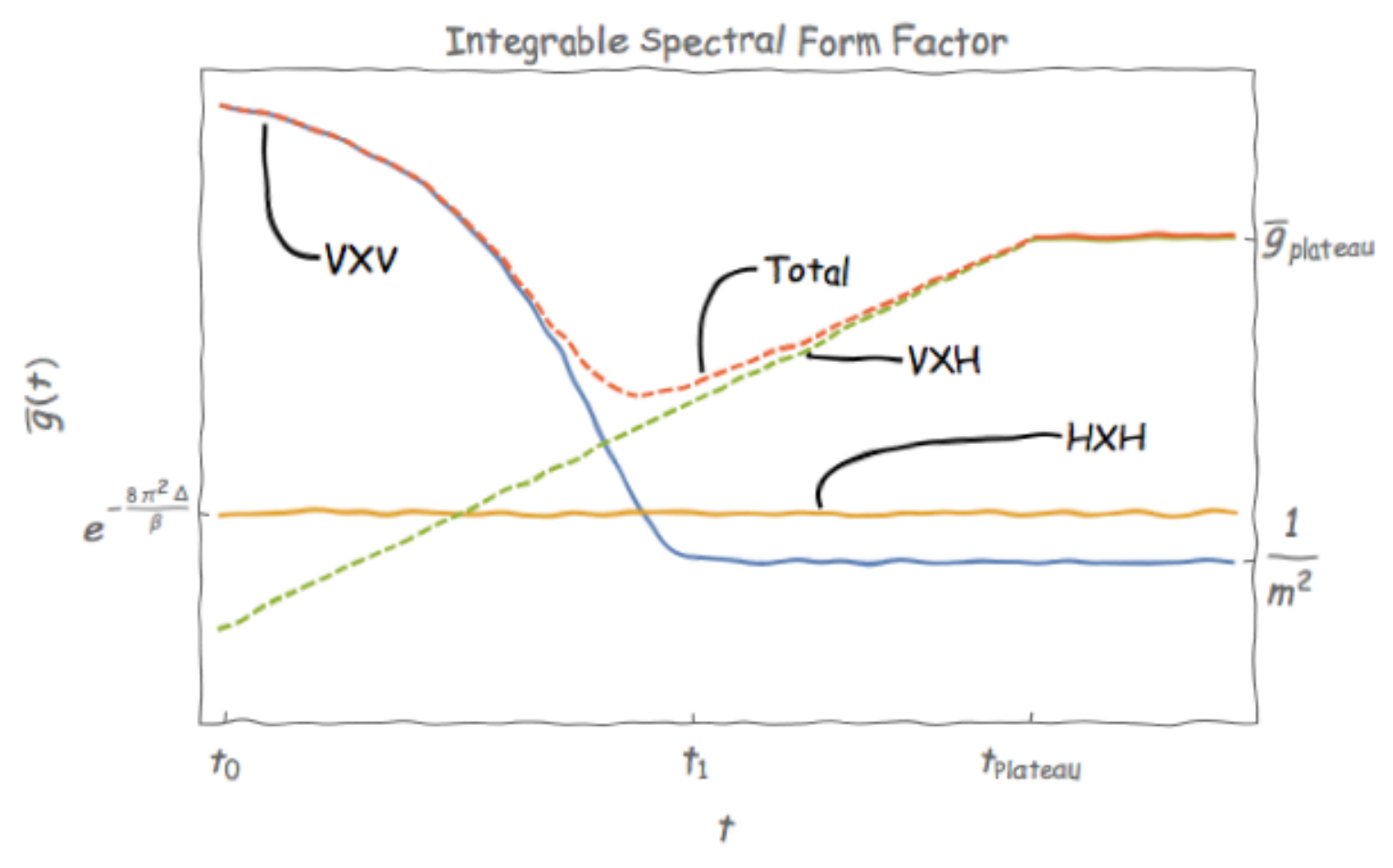}
\caption{{\bf Left:} A schematic form for the SFF in a typical chaotic theory. {\bf Right:} A plot of the purported large $m$ behavior of the SFF in our integrable models. The solid lines correspond to bounds or asymptotic behavior derived in the text; the dashed lines correspond to conjectured large $m$ behavior for the $V\times H$ component.}
\label{fig:Conjecture}
\end{center}
\end{figure}

\item{$\mathbf{V\times H}$:} In some sense this is the most interesting contribution to the partition function. Initially this contribution is zero, due to the orthogonality of different rows of $S$. At late times, however, the above arguments imply that the $V\times H$ contribution must make up the dominant contribution to the plateau.

It would be interesting to understand the growth of this factor in detail. For instance, \cite{Balasubramanian:2016ids} find logarithmic growth in the SFF for the D1-D5 system at the orbifold point, while it is conjectured that generic, non-integrable theories have linear growth. Though we cannot compute the nature of this ramp, we conjecture that the contribution is small initially, leading to the qualitative \textit{dip-ramp-plateau} form for the SFF. 

The conjectured large $m$ behavior of these three contributions are summarized in  Figure \ref{fig:Conjecture}.

If this conjecture for the $V\times H$ contribution is correct, then the constructions described here represent a novel example of functions exhibiting this \textit{dip-ramp-plateau} structure. Furthermore, this basic approach can likely be extended beyond the minimal models to other parametric families of large dimension representations of $\sltz$. 
\end{itemize}

\begin{figure}
  \centering
  \raisebox{0pt}{
    \includegraphics[width=0.45\linewidth]{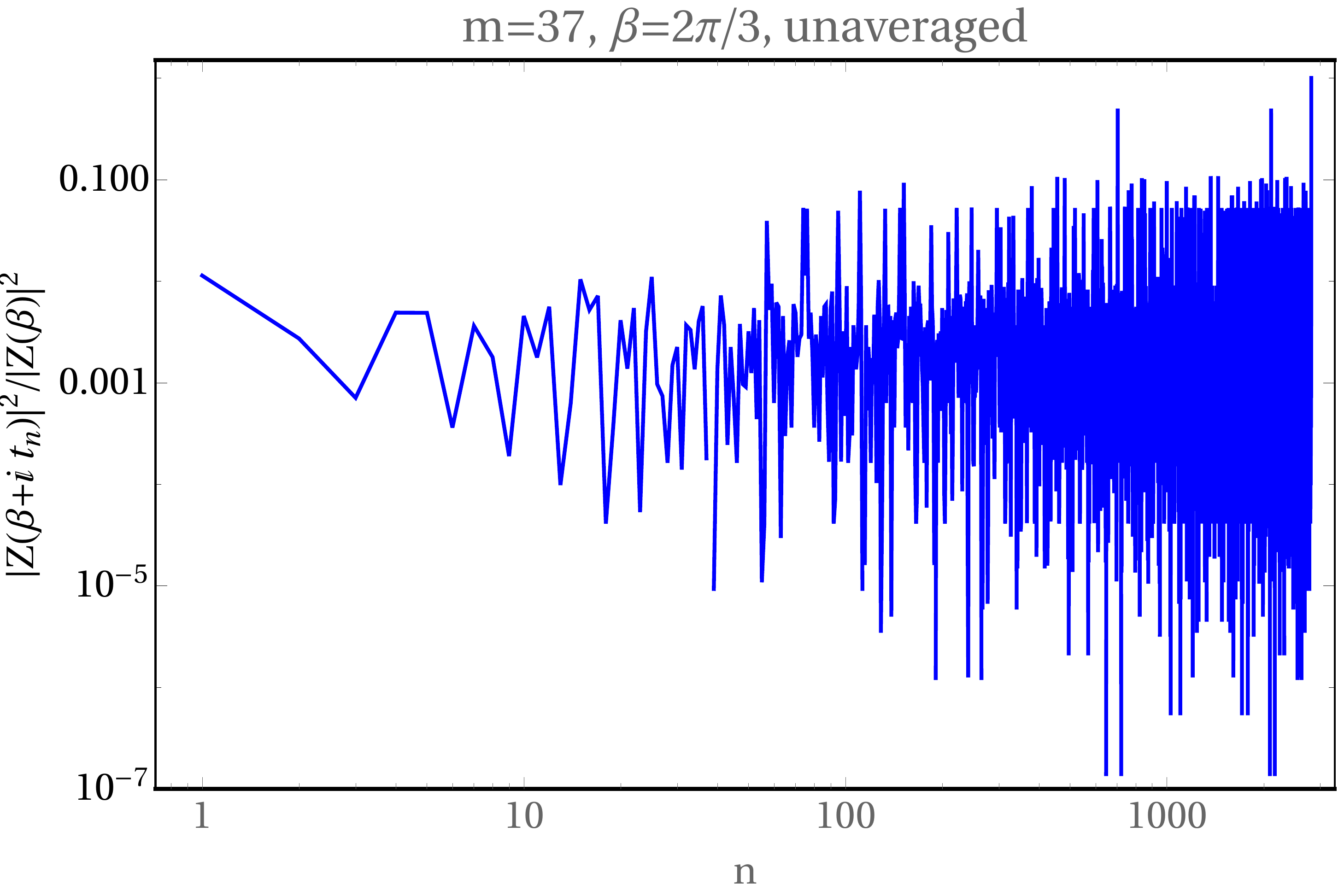}
   } ~~  
  \includegraphics[width=0.48\linewidth]{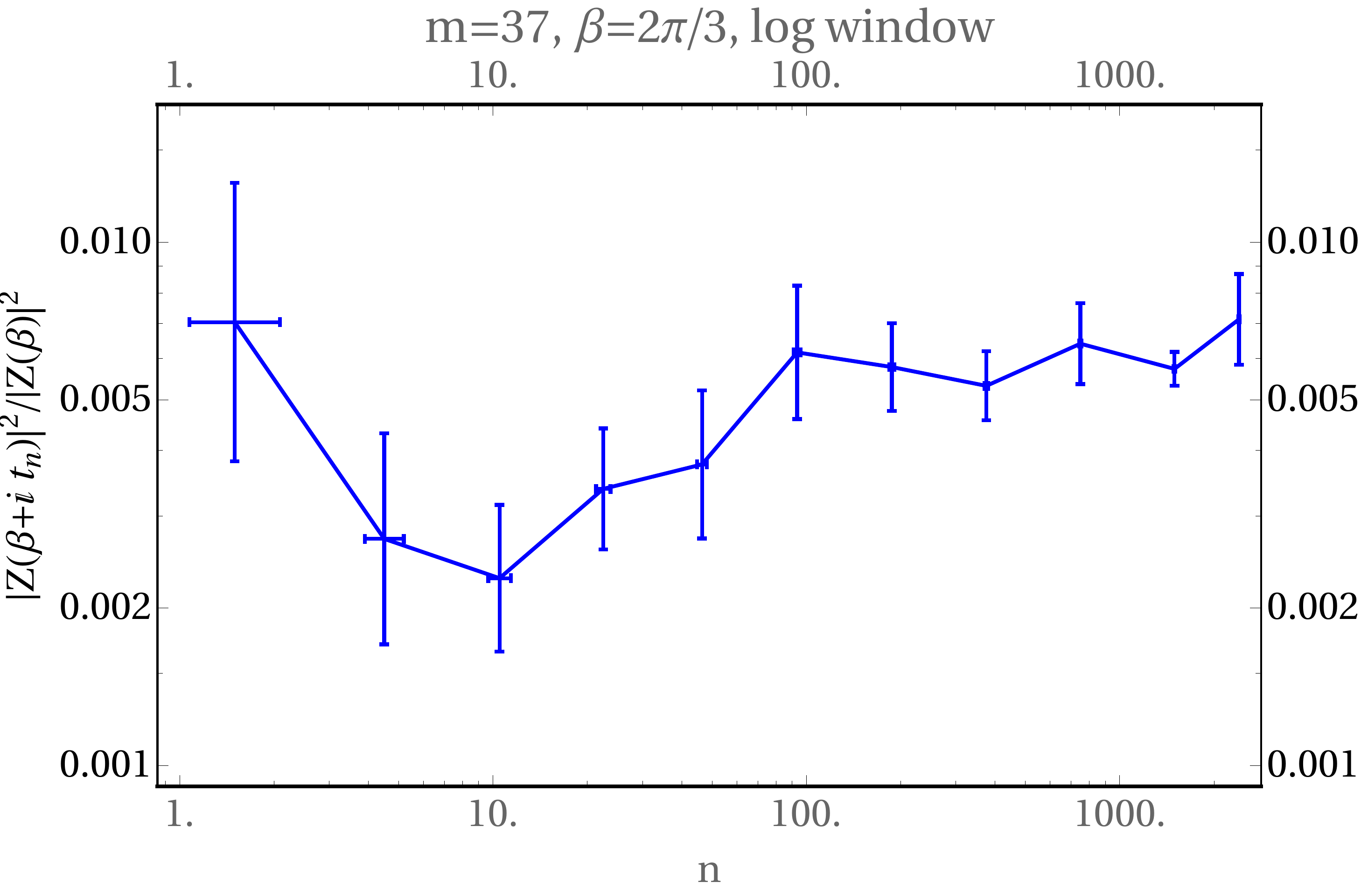}
  \caption{
    \label{fig:m_10_gappy_exact_disc}
    {\bf Left:} The spectral form factor of the $m=37$, $k=1$ model at integer-spaced times $t_n = 2\pi n$ without averaging. 
    {\bf Right:} The same plot averaging over a log window of $t$. The error bar of each point is the standard deviation of the mean.  At these values of $m$, which are the largest we are able to compute exactly, it is still unclear what the large $m$ behavior of the function approaches.
  }
\end{figure}

Using the Bantay-Gannon construction, we can construct vectors for partition functions with central charge $c = 24k + \OO(1)$ and gap $\Delta_* \approx \frac{c}{24}$, and study their SFFs.
The SFFs of these ``large $c$ large gap'' partition functions share many similarities with those of the minimal models discussed in Sec.~\ref{sec:minimal_models}. See  Fig.~\ref{fig:m_10_gappy_exact_disc} for an example of a small $m$ ($m=37)$ SFF.

%Fig.~\ref{fig:m_10_gappy_exact_cont}\footnote{The temperature we choose in this plot is lower than the previous example because the convergence of the $q$-expansion as $|t|$ grows gets worse more rapidly at higher temperature, and the ability to compute more terms in the $q$-expansion is limited by the machine precision.} 
%and Fig.~\ref{fig:m_10_gappy_exact_disc} for an example of small ($m=10$ and $m=37$) SFF. The mini-recurrence is sharper due to the large gap and the SFF is strongly dominated by the integral spaced time $t_n = 2\pi n$. If we focus on these values, and average properly with a parametrically small time window, we get a semiclassical decay at early time, and a ramp followed by a plateau at late time.%, when $m$ is large enough for the ultimate recurrence to come sufficiently late to allow scrambling.

Ideally, we would be able to efficiently take $m$ to be much larger.  However, the computational difficulty grows much more quickly with $m$ than for the minimal models, where we were able to take $m \sim 10^4$.  We will therefore have to make some approximations. There is an obstruction to efficient numeric computation of (\ref{matrixevolution}), which is that we have to compute the $t=0$ values of all the characters. We are only able to analyze the $V\times V$ piece of (\ref{matrixevolution}). As we have seen, for sufficiently large $m$ (in particular the regime where $m \gg e^c$), this is not the dominant contribution except at early times. However, in other parametric regimes even where $m$ is large (but much less than $e^c$), this is the dominant contribution at all times.

%To deal with this, we will need to approximate the characters on the right hand side of (\ref{matrixevolution}). The leading order at high temperature $\b\rightarrow 0$ is from the vacuum state. We will first study the contribution purely from the vacuum piece and analyze its behavior. The behavior of this piece will be interesting in its own right. We will also do a better approximation which will be to estimate the nonvacuum characters, as described below.

\subsection{Vacuum approximation}

\begin{figure}
  \centering
  \includegraphics[width=0.45\linewidth]{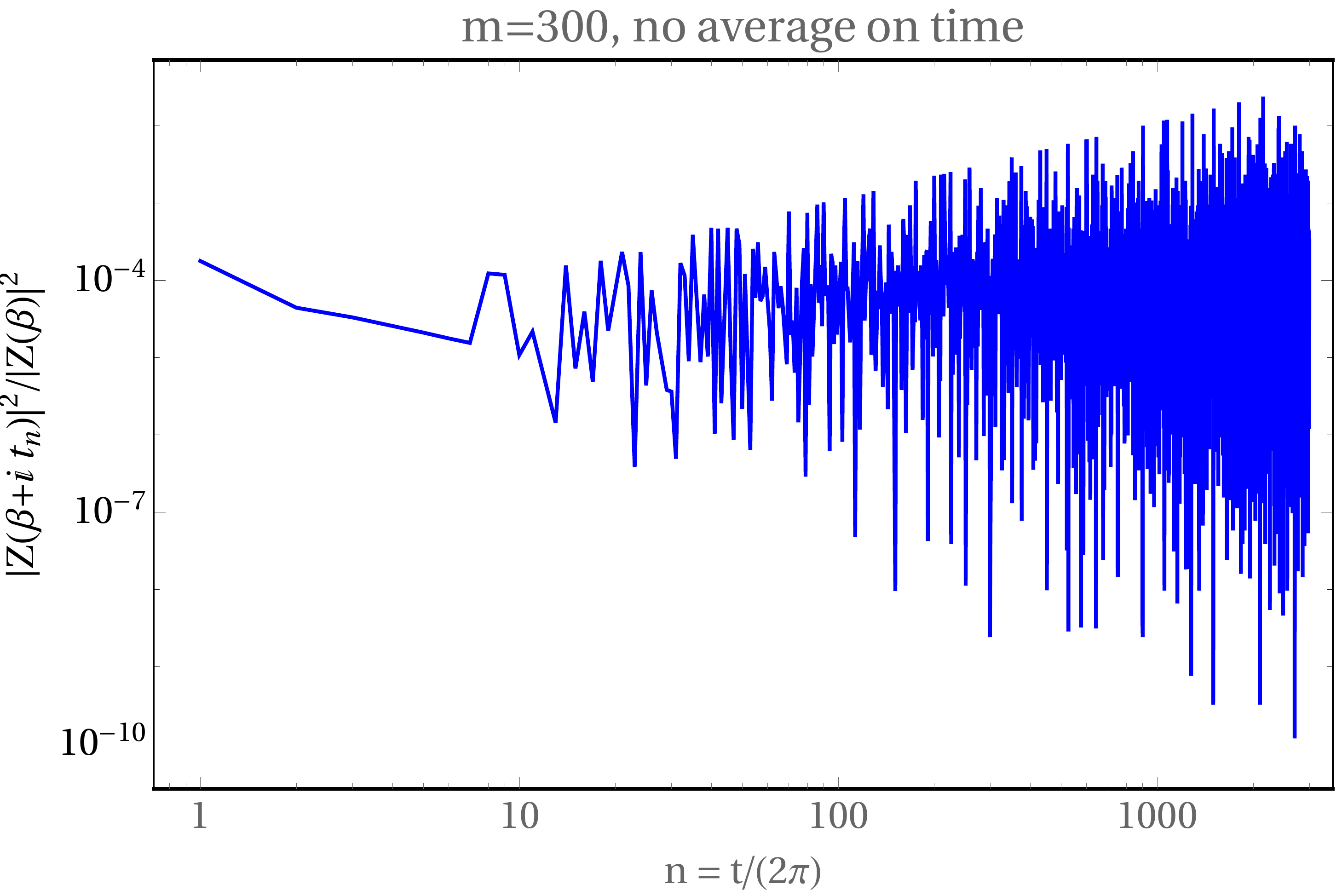} ~~
  \includegraphics[width=0.45\linewidth]{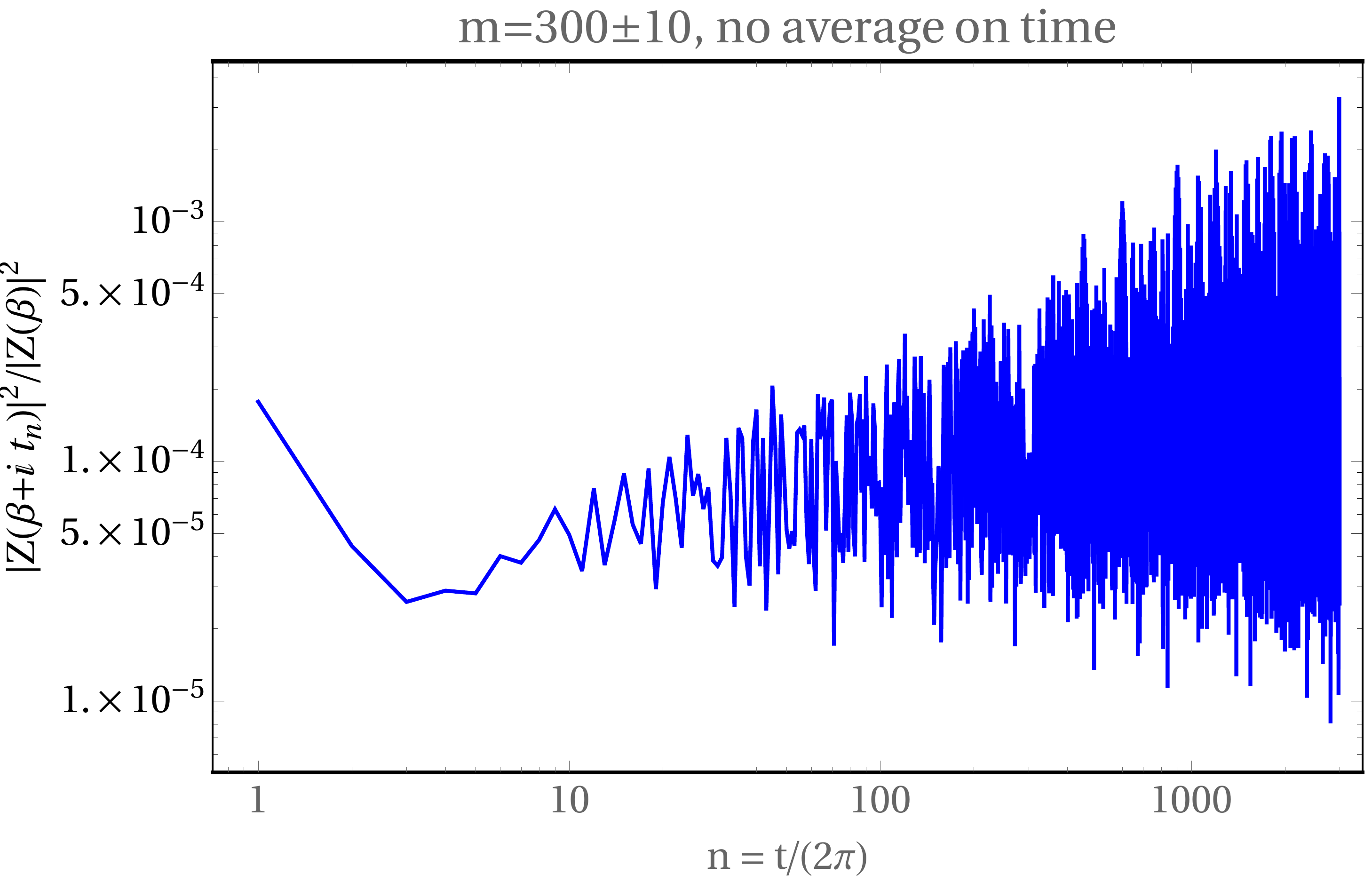} \\
  \includegraphics[width=0.48\linewidth]{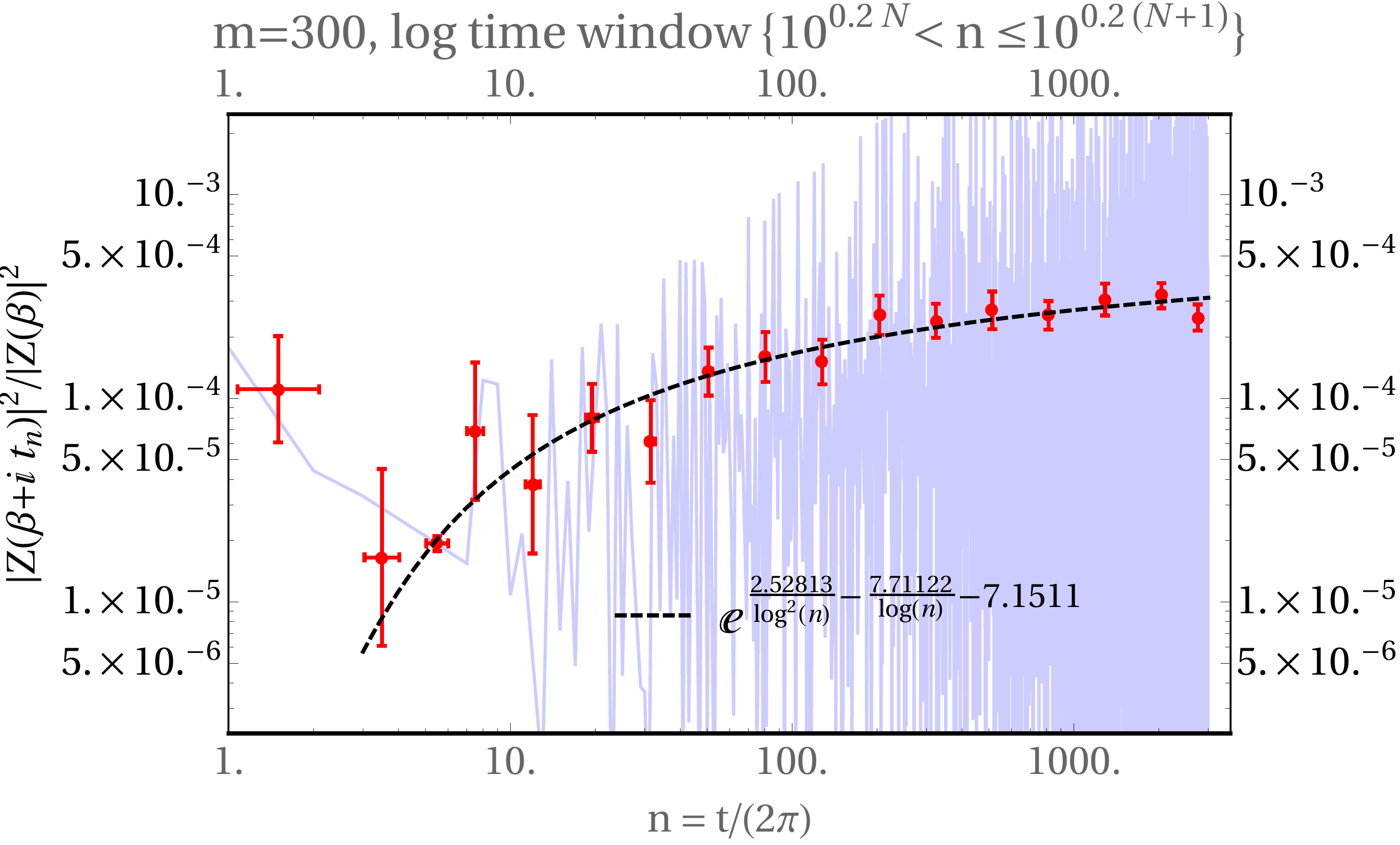} ~~
  \includegraphics[width=0.48\linewidth]{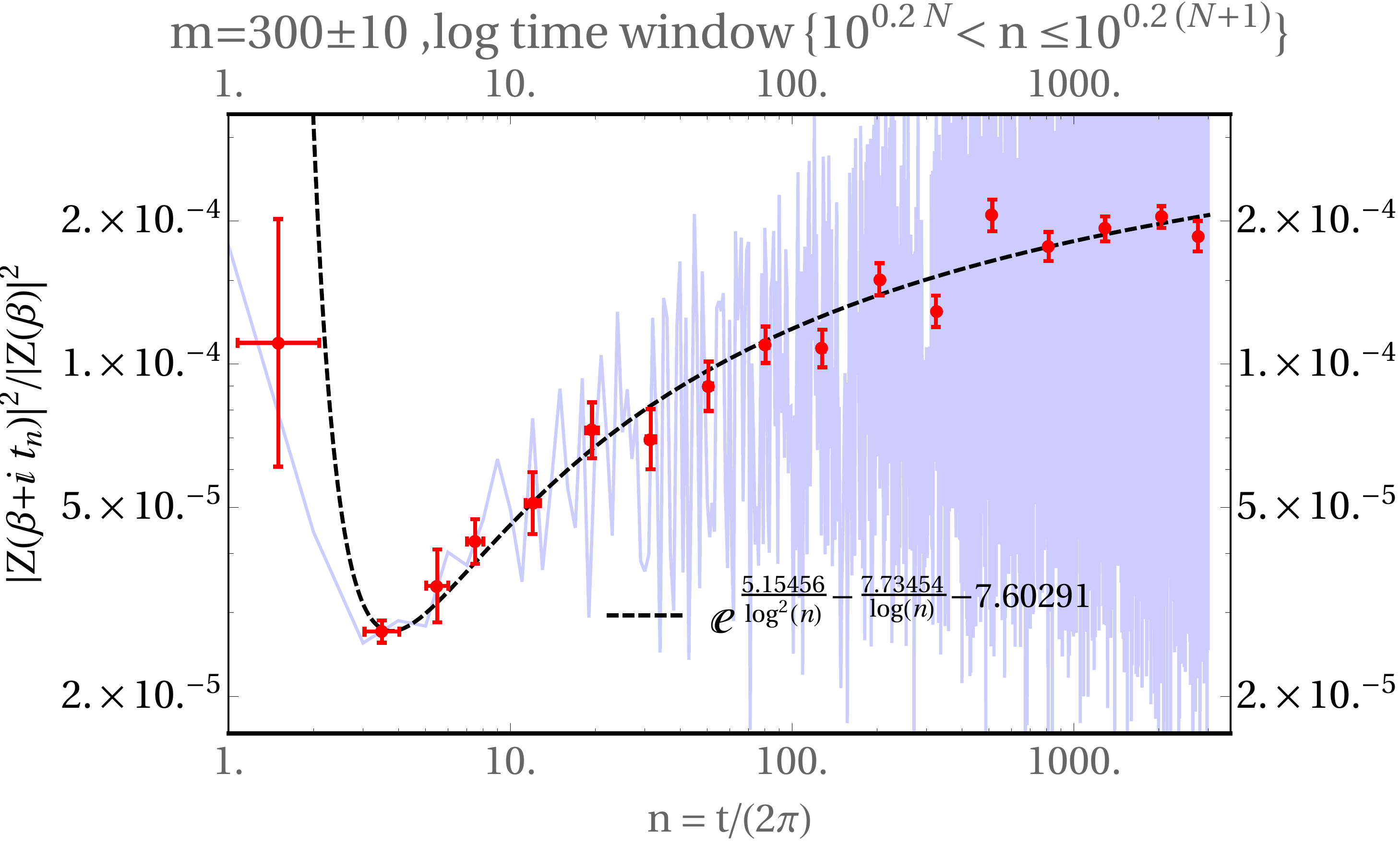} 
  \caption{
    \label{fig:st2ns00}
    The $V\times V$ portion of the SFF of the $m=300$ model. 
    {\bf Upper left:} The $V \times V$ SFF before any averaging. 
    {\bf Upper right:} The $V \times V$ SFF after averaging over a small window of $290 \leq m\leq 310$. 
    {\bf Lower left:} The $V \times V$  SFF of a single model $m=300$, averaged over a log window of time. 
    {\bf Lower right:} The $V \times V$ SFF with both $m$ and time averaged. 
    The data and error bars of the lower plots are the mean and the standard deviation of the mean over each time window. The log-log plot of the $V \times V$ SFF fits well to inverse powers of $\log(n)$, at intermediate and late time. 
  }
\end{figure}

If either the gap is large or the temperature is low, $(\frac{k}{\b} \rightarrow \infty )$, the vacuum state in the zeroth component dominates the vector $\chi^\m \(\frac{4\pi^2}{\b}\) \approx \delta_0^\m e^{-\frac{4\pi^2 \Delta_0}{\b}}$. If we plug this approximation into (\ref{matrixevolution}) and normalize by the $(t=0)$ value, we get the dominant piece of the normalized SFF as just the $(0,0)$ component of the modular transformation matrix $\CM (ST^{2n}S)$
\begin{align}
  \hat g(\b, t_n) \equiv \abs{ \frac{  Z(\b + i t_n) }{ Z(\b)} }^2 = \abs{ \CM (ST^{2n}S)_{00} }^2
  + \text{ (exponentially suppressed terms). }
\end{align}
An example of such an SFF is shown in Fig.~\ref{fig:st2ns00}. The plot is highly oscillatory at late times so averaging over a window of $n$ and $m$ helps in finding the ramp and plateau.

\begin{figure}
  \centering
  \includegraphics[width=0.48\linewidth]{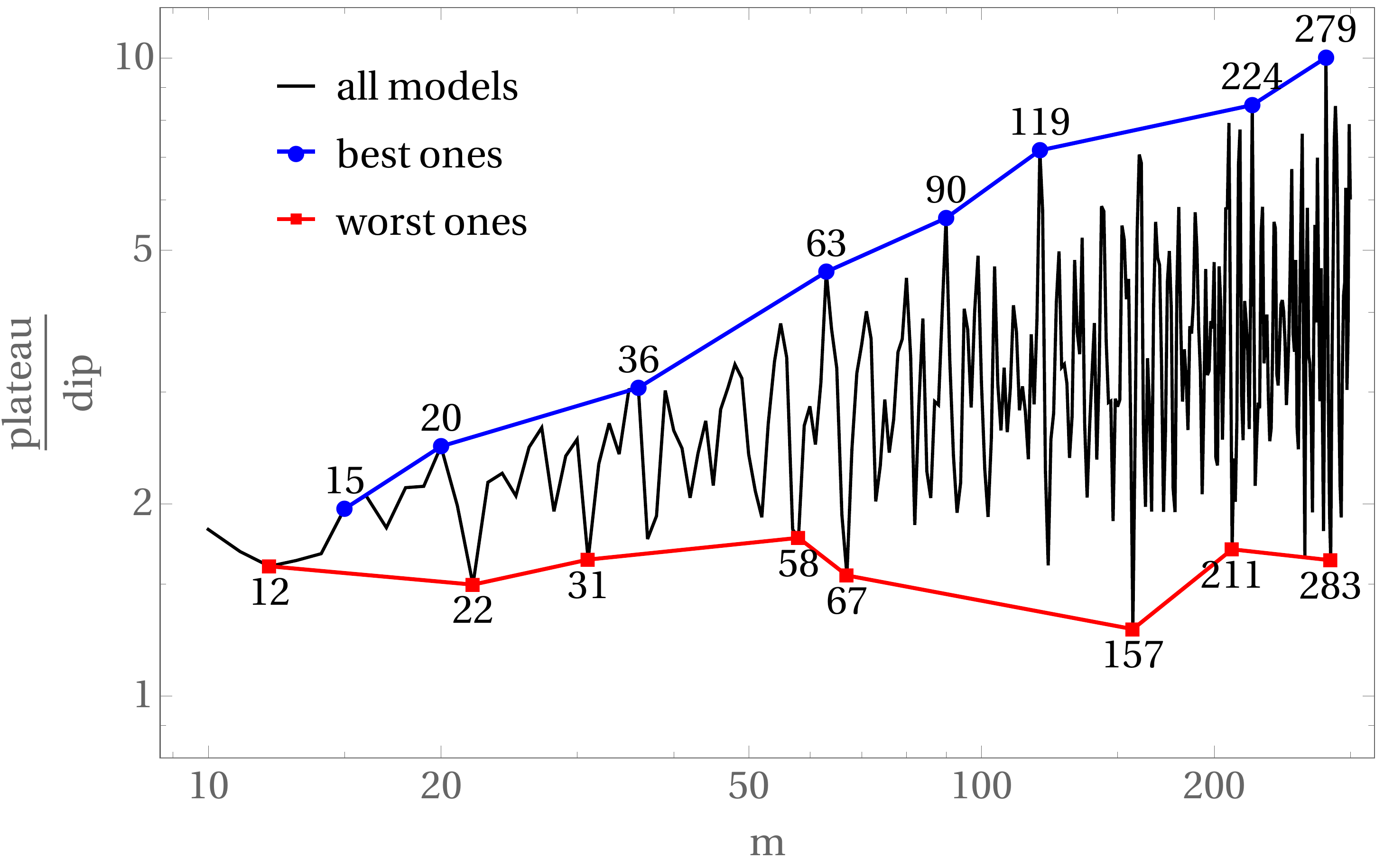}~~
  \raisebox{-30px}{
    \includegraphics[width=0.48\linewidth]{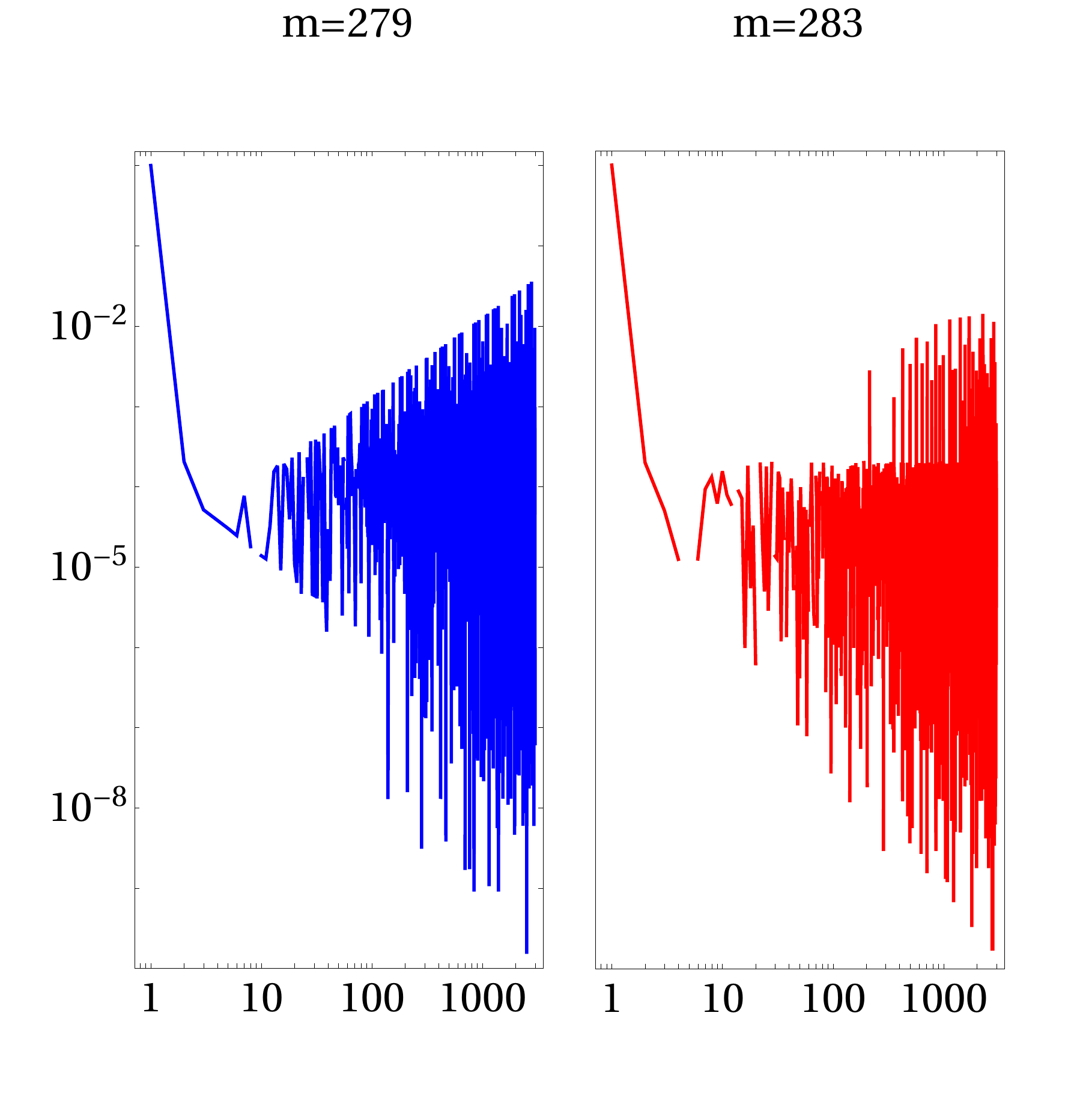}
  }
  \caption{
    \label{fig:st2ns00_ratio}
    {\bf Left:} The behavior of the SFF's plateau height over dip height ratio as a function of $m$. The plateau is computed using (\ref{eq:ethan_average}) and the dip is the mean of SFF between $n=1$ and $10$ without any average over $m$.
    The best and worst models are labeled by blue circles and red squares. The ratio grows with $m$ for the best models while the worst models seem not to grow with $m$.
    {\bf Right:} Comparing two example SFFs with the large and small plateau-to-dip ratio. Their $m$ indices are close but their behaviors are drastically different at late time. The best model ($m=279$) occasionally gets increasingly constructive interference that helps the SFF to grow, while the worst model ($m=283$) does not have constructive interference.
  }
\end{figure}

We average the SFF
\begin{align}
  \bar{g}(\beta)&=\frac{1}{N}\sum_{n=1}^{N}g(\beta,t_{n})~,
  \label{eq:ethan_average}
\end{align}
to compute the late time plateau. Both the dip and the plateau for the $V\times V$ contribution to the SFF are down by an overall factor of $\frac{1}{m^2}$. However, their relative dominance is highly $m$-dependent, and depicted in Fig. \ref{fig:st2ns00_ratio}, obscuring any sharp $m$-independent ramp statement. 

{\bf Approximating the non-vacuum states}

The character $\chi$, as a vector-valued modular form, can be computed by summing over the images of the vacuum state under modular transformation. This convergent infinite sum is called a ``Rademacher sum'', reviewed in Appendix \ref{sec:radsum}. The sum is dominated by its first term, i.e. the image under $S$ transformation. Thus we can use this first term as a good approximation of the non-vacuum states. Throughout the numerically computable regime where $m$ is a few hundred, the non-vacuum piece is negligibly small compare to the vacuum contribution, for the gapped models. In the regime $m \gtrsim e^{\frac{2\pi^{2}k}{\beta}}$ this non-vacuum piece becomes important.

\section{Discussion}
\label{sec:discuss}

In this paper, we proposed a family of partition functions for 2d CFTs that we conjecture to be dual to pure AdS$_3$ gravity. Our construction, based on techniques introduced in \cite{Bantay:2005vk} satisfies basic consistency requirements for 2d CFT partition functions, namely modular invariance and positivity. In the limit of large $c$ and large $\sltz$ representation, we argue that our partition functions exhibit random-matrix-like behavior of the spectral form factor. This random-matrix-like behavior is conjectured to be a universal feature in CFTs dual to semiclassical Einstein gravity, and is missing in other proposals for pure 3d gravity partition functions such as \cite{Witten:2007kt}. Furthermore we have demonstrated that an avatar of this random-matrix-like behavior is exhibited even in the minimal models at large $m$, where we have analytically derived the dip, linear ramp, and plateau. 

The most important question we have not addressed, of course, is whether there exist CFTs with the partition functions we proposed. Since our construction of the partition functions was based off of the minimal model partition functions (or more generally, that of any RCFT), it would be interesting if there were some method of constructing a full CFT based off of the existence of the minimal models. Another possible avenue would be using the conformal bootstrap to construct (or disprove) our proposed partition functions at large $c$ and large $\sltz$ representation. 

Of course minimal models are complete theories, with known correlation functions, and these can be studied in more detail at large $m$. It would be interesting to see if they exhibit chaotic features beyond simply the SFF, such as behavior of the out of time ordered four point functions. In practice, computing these correlation functions is computationally expensive, especially for correlators with many operators in their OPE. Nevertheless, code is available publicly that can in principle compute all of the quantities involved.  Mathematica code is provided in \cite{Chen:2017yze} to efficiently compute Virasoro conformal blocks, and in \cite{Esterlis:2016psv} to compute the OPE coefficients of primary operators in minimal models.  Putting these two pieces together constructs arbitrary four-point functions in the minimal models.  

Another interesting question would be to improve on our partition functions. There are two non-generic features our partition functions have. One is the presence of conserved currents at $\mathcal{O}(c)$ above the vacuum. This can be seen from our construction of the vector $\chi^{\mu}$ -- the vacuum component of this vector, $\chi^{0}$ contains new Virasoro primary states at $\mathcal{O}(q^0)$ and above, which when multiplied with $\overline{\chi^0}$ correspond to currents in the theory. The second (related) non-generic feature is that we have infinitely many Virasoro primaries whose conformal weights differ from each other by an exact integer. Again, these simply come from the fact that in any component of our vectors $\chi^{\mu}$, there are multiple Virasoro primaries. It would be interesting to construct a large $c$ partition function without these features. However, in our construction, in the limit of large representation size, both currents and primaries that are integer-separated from other primaries are measure zero, which is key for the random-matrix-like behavior of the spectral form factor to emerge. It would also be interesting to investigate whether partition functions with a larger gap exist. Similarly it would be interesting to relax our condition of ``pure" gravity. For instance, it is possible that the SFF changes qualitatively when the low-lying density of states changes from sub-Hagedorn to Hagedorn.

Finally, we briefly comment on connections to other work. In \cite{Harvey:2018rdc}, a method of generating new vectors that transform as a vector-valued modular form based on any RCFT by using Hecke operators was proposed. Given a $d$-dimensional RCFT, by acting on the $d$ characters with the Hecke operator $T_p$, one can construct a new set of $d$ vectors that transform in a (possibly different) $d$-dimensional representation of $\sltz$, where the pole in the vector gets multiplied by $p$. Moreover, for certain choices of $p$, positivity is guaranteed. Unfortunately, for generic seed RCFT, we cannot construct a large $c$ CFT that has all of the required Virasoro vacuum descendants using this technology; nonetheless for specific RCFTs with small number of representations (e.g. the Ising model), it would be interesting to see the connection to \cite{Bantay:2005vk}. One other interesting direction would be to use techniques of \cite{Aharony:2018bad} and study the $T\bar T$ deformed partition function of our theories\footnote{We thank H. Verlinde for pointing out this question to us.}.

\ack{We thank Guy Gur-Ari, Shamit Kachru, Jared Kaplan, Ami Katz, Chris Laumann, Anatoli Polkovnikov, Brandon Rayhaun, Douglas Stanford, Herman Verlinde, and Yifan Wang for helpful discussions.
 NB is supported by a Stanford Graduate Fellowship and an NSF Graduate Fellowship. ALF and YX were supported in part by the US Department of Energy Office of Science under
Award  Number  DE-SC-DE-SC0015845.  ALF  was also supported in part by a Sloan Foundation Fellowship, and ED and ALF are supported in part by the Simons Collaboration Grant on the Non-Perturbative Bootstrap.}

\begin{appendices}
%\appendix

\section{Analytic Derivation of Dip and Ramp in Minimal Models}
\label{app:CambridgeAnalytica}

The projected partition function for the diagonal minimal models is given by equation (\ref{eq:MinModelZ}), reproduced here:
\be
Z(\tau) = \frac{\vartheta_3(0,e^{\frac{i\pi \tau}{m(m+1)}})\vartheta_3(0,e^{i\pi m(m+1)\tau})-\vartheta_3(0,e^{\frac{i\pi(m+1) \tau}{m}})\vartheta_3(0,e^{\frac{i\pi m\tau}{m+1}})}{2\eta(\tau)^2}.
\ee
In the large $m$ limit, the interesting behavior of this partition function is given by the first $\vartheta$ factor:
\be
g(t,\beta) = |\vartheta_3(0, e^{\frac{i\pi \tau}{m(m+1)}})|^2 ,
\label{eq:minimalapprox}
\ee
with $\tau = \frac{i\beta}{2\pi} + t$.
The initial $\frac{1}t$ decay at fixed $t$ and large $m$ can be seen from an $S$ transformation of the $\vartheta_3$ term.

\begin{figure}[t!]
\begin{center}
\includegraphics[width=0.8\textwidth]{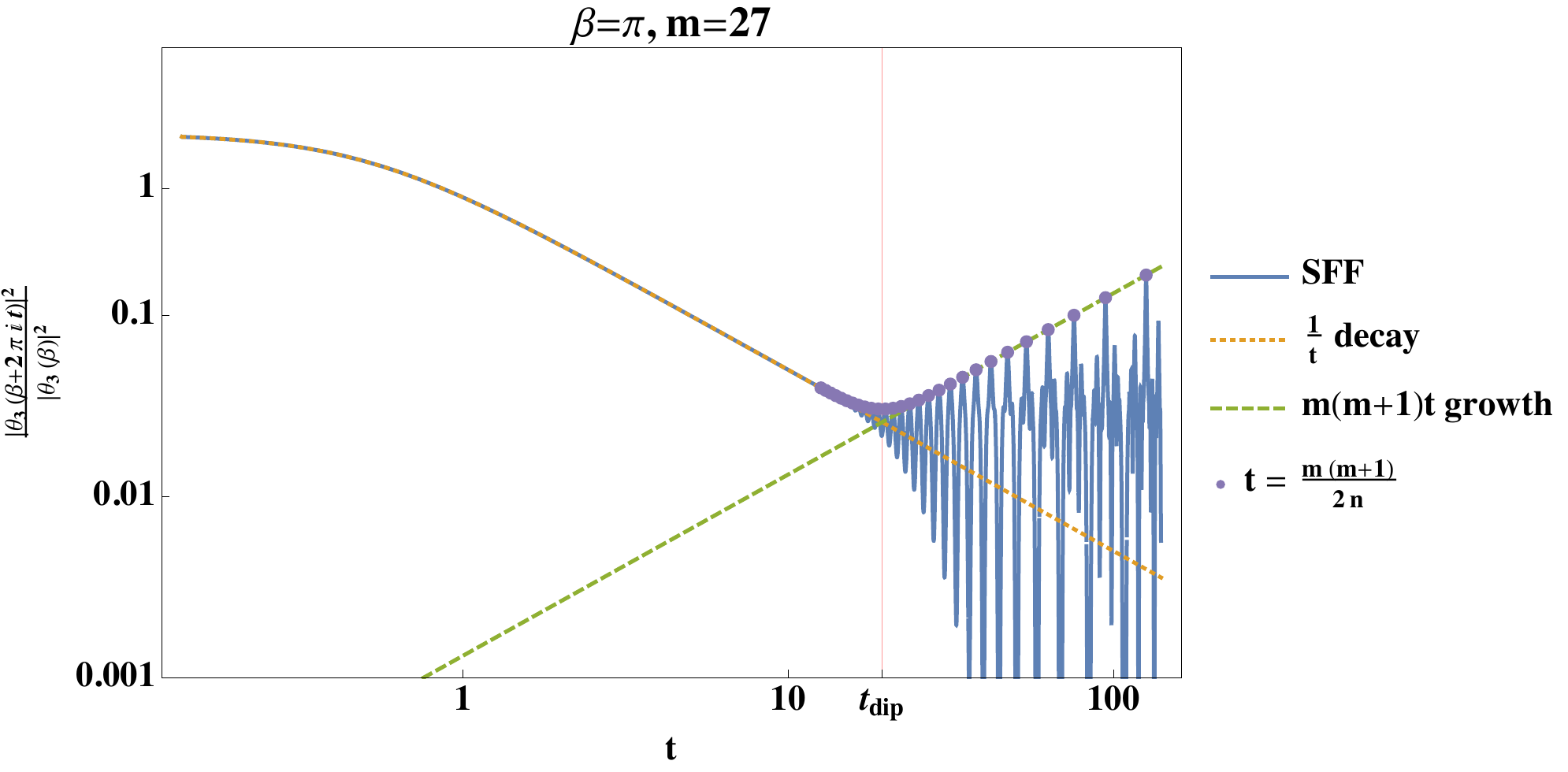}
\caption{Plot of the behavior of $\vartheta$ in (\ref{eq:minimalapprox}).  The peaks at times beyond $t_{\rm dip}$ occur at $\frac{t}{m(m+1)} = \frac{1}{2n}$ for integer $n$, and are highlighted with purple points in the plot.}
\label{fig:ThetaPlot}
\end{center}
\end{figure}

Next we will demonstrate the linear growth of the SFF at times $t \sim \mathcal{O}(m)$. Moreover, we can see numerically that the dominant contributions to (\ref{eq:minimalapprox})  come from $\frac{t}{m(m+1)} = \frac{1}{2n}$ for some integer $n$ The corresponding peaks are highlighted with purple points in Fig. \ref{fig:ThetaPlot}.

Recall that (up to a phase), $\vartheta_3(0, e^{i \pi \tau})$ is a weight $\frac12$ modular form under the $\sltz$ congruence subgroup $\Gamma_{\theta}$, which is generated $S$ and $T^2$. Suppose $t = \frac{m(m+1)}{2n}$ for some integer $n$. Let's define
\begin{align}
\tilde \tau &= \frac{\tau}{m(m+1)} = \frac{i\beta}{2\pi m(m+1)} + \frac{1}{2n} \nn\\
\tilde \tau' &= \frac{\tilde \tau}{-2n \tilde \tau + 1} = \frac{i m(m+1)\pi}{2n^2\beta} - \frac{1}{2n}
\end{align}

We can rewrite (\ref{eq:minimalapprox}) as 
\begin{align}
g(t, \beta) &= |\vartheta_3(0, e^{i \pi \tilde \tau})|^2 \nn\\
&= |(-2n \tilde\tau + 1)^{-\frac12} \vartheta_3(0,e^{i\pi \tilde \tau'})|^2 \nn\\
&= \frac{t}{\beta}|\vartheta_3(0, e^{i\pi \tilde \tau'})|^2
\label{eq:omgramp}
\end{align}

When $\tilde\tau'$ is close to $i \infty$, the $\vartheta_3(0,e^{i \pi \tilde \tau'})$ approaches $1$, and the only $t$ dependence is from the prefactor, giving us a linear growth. This is a good approximation when 
\be
m \gg n \sqrt\beta.
\ee
Moreover, sampling only the times $t = \frac{m(m+1)}{2n}$ (i.e. the dots in Fig \ref{fig:ThetaPlot}) is a good approximation when the difference between successive times is at most $\mathcal O(1)$. This is the case when
\be
m \lesssim n,
\ee
meaning $n$ grows at least linearly with $m$. Thus we can trust (\ref{eq:omgramp}) when $t \sim \mathcal{O}(m)$ at small $\beta$. 

\section{Details of the Bantay-Gannon Construction}
\label{sec:bgapp}

In \cite{Bantay:2005vk, Bantay:2007zz, Bantay:2011}, the space of vector-valued modular forms was studied. Consider a finite-dimensional representation of $\sltz$, $$\rho\,:\,\sltz \rightarrow GL(r,\mathbb{C}),$$ which is generated by the two matrices $S$ and $T$. An arbitrary matrix for the $\sltz$ element $\tau \rightarrow \frac{a \tau + b}{c\tau+d}$ is denoted by $\rep abcd$. We focus on the representations that diagonalize the $T$ matrix with the form
\begin{equation}
T = \exp \left( 2 \pi i \Lambda \right),~~~\text{where}~~ \Lambda = \text{diag}(\lambda_1,\,\lambda_2,\,\cdots, \lambda_r)~.
\end{equation}
The CFT interpretation of these elements is
\be
 \lambda_i = h_i - \frac{c}{24} ~\text{(mod}~1)
 \ee
 where $h_i$ is the conformal weight of the $i^{\text{th}}$ primary operator. This fixes the vector-valued modular form $\X$ to be of the form
\be
\X = \sum_{n\in \mathbb{Z}}\X[n]\, q^{n + \lambda_{\mu}}.
\ee
The exponent matrix $\Lambda$ is unique up to integral shift. Once the
matrix $\Lambda$ is fixed, the form is uniquely determined by the polar part
\begin{equation}
  \label{eq:polar}
  \mathcal{P}\X \equiv q^{-\Lambda}\sum_{n<0}\X[n]\,q^{n}~.
\end{equation}
Consistent choices of $\Lambda$ have to satisfy
\begin{equation}
  \label{eq:consistent-lambda}
  \tr(\Lambda) = \frac{5d}{12} + \frac{1}{2} \tr(S) +
  \frac{2}{3\sqrt{3}}{\rm Re}\left( e^{-\frac{\pi i}{6}}\tr(U) \right)~,
\end{equation}
where $U = \rep{0}{-1}{1}{-1}$ .

The space $\mathcal{M}(\rho)$ of all vector-valued modular forms of representation $\rho$ is a polynomial algebra $\mathbb{C}[J]$ where $J(\tau)$ is the unique $\sltz$-invariant modular function with pole at $\tau=i\infty$ that goes as
\begin{equation}
  \label{eq:j-function}
  J(\tau) = q^{-1} + 196884 \, q + \cdots.
\end{equation}
We take functions
$\X^{(\xi;n)}$, that have only one order-$n$ pole at component $\xi$
to be a canonical basis of  $\mathcal{M}(q)$:
\begin{equation}
  \label{eq:canonical-basis}
  \left[ \mathcal{P}\X^{(\xi;n)}(q) \right]_\eta = q^{-n}\delta_{\xi \eta}~.
\end{equation}
Higher order basis are determined by recursion relation
\begin{equation}
  \label{eq:recursion-relation}
  \X^{(\xi;m+1)} = J(\tau) \X^{(\xi;m)}-\sum_{n=1}^{m-1} J_n \X^{(\xi;m-n)} -
  \sum_\eta \,\left(\X^{(\xi;m)}[0]\right)_\eta \, \X^{(\eta;1)},
\end{equation}
where $J_n$ is the coefficient of $q^n$ in $J(\tau)$'s expansion.

As an example, in the next two subsections, we will explicitly write down the basis vectors $\X^{(n;1)}(\tau)$ for the Ising model and the Tricritical Ising model. By taking linear combinations of these characters, combined with multiplying by arbitrary powers of $J(\tau)$, we can get any vector $\X^{(n;m)}(\tau)$.

\subsection{Example: Ising Model}

The basis vectors $\X^{(n;1)}$ for the Ising model ($n=1, 2, 3$) were worked out in Section 7 of \cite{Bantay:2005vk}; we reproduce the results below. The $S$ and $T$ matrices for the Ising model are given below:
\begin{align}
S &= \begin{pmatrix} \frac12 & \frac12 & \frac1{\sqrt{2}} \\ \frac12 & \frac12 & -\frac1{\sqrt{2}} \\ \frac1{\sqrt{2}} & -\frac1{\sqrt{2}} & 0 \end{pmatrix} \nonumber \\
T &= \begin{pmatrix} e^{2\pi i \left(\frac{47}{48}\right)} & 0 & 0 \\ 0 & e^{2\pi i \left(\frac{23}{48}\right)} & 0 \\ 0 & 0 & e^{2\pi i \left(\frac{1}{24}\right)} \end{pmatrix}
\label{eq:stising}
\end{align}
As in Section 7 of \cite{Bantay:2005vk}, define
\begin{align}
f(\tau) &= q^{-\frac{1}{48}}\prod_{n=0}^{\infty} (1+q^{n+\frac12}) \nonumber \\
f_1(\tau) &= q^{-\frac{1}{48}}\prod_{n=0}^{\infty} (1-q^{n+\frac12}) \nonumber \\
f_2(\tau) &= \sqrt{2} q^{\frac{1}{24}} \prod_{n=1}^{\infty} (1+q^n)
\end{align}
then
\begin{align}
\X^{(1;1)} &= \frac12 \begin{pmatrix} f+f_1 \\ f-f_1 \\ \sqrt{2} f_2 \end{pmatrix}
\end{align}%
\begin{align}
\X^{(2;1)} &= \frac12 \begin{pmatrix} f^{25}-f_1^{25}-25f - 25f_1\\ f^{25}+f_1^{25}-25f+25f_1 \\ -\sqrt{2} f_2(25+f_2^{24}) \end{pmatrix}
\end{align}%\nonumber \\
\begin{align}
\X^{(3;1)} &= \begin{pmatrix} 8f^{17}f_1^8 - 8 f^{24} f_1 - 128f + \frac{f_2^7}{\sqrt{2}}(f^{39}-f_1^{39}-16f^{15}-32f_1^{15}) \\ 8f^{17}f_1^8 + 8 f^{24} f_1 - 128f - \frac{f_2^7}{\sqrt{2}}(f^{39}+f_1^{39}-16f^{15}+32f_1^{15}) \\ f^{15}f_1^7(f^{24}-16)-8\sqrt{2}f_2f^{24} \end{pmatrix}
\end{align}

\subsection{Example: Tricritical Ising Model}

Here we provide the basis vectors $\X^{(n;1)}$ for the Tricritical Ising model $(n=1, 2, \ldots, 6)$. The $S$ and $T$ matrices for the Tricritical Ising model are given below:
\begin{align}
S &= \left(
\begin{array}{cccccc}
 \frac{s_2}{\sqrt{5}} & \sqrt{\frac{2}{5}} s_2 & \frac{s_1}{\sqrt{5}} &
   \frac{s_1}{\sqrt{5}} & \sqrt{\frac{2}{5}} s_1 & \frac{s_2}{\sqrt{5}}
   \\
 \sqrt{\frac{2}{5}} s_2 & 0 & -\sqrt{\frac{2}{5}} s_1 & \sqrt{\frac{2}{5}}
   s_1 & 0 & -\sqrt{\frac{2}{5}} s_2 \\
 \frac{s_1}{\sqrt{5}} & -\sqrt{\frac{2}{5}} s_1 & -\frac{s_2}{\sqrt{5}}
   & -\frac{s_2}{\sqrt{5}} & \sqrt{\frac{2}{5}} s_2 &
   \frac{s_1}{\sqrt{5}} \\
 \frac{s_1}{\sqrt{5}} & \sqrt{\frac{2}{5}} s_1 & -\frac{s_2}{\sqrt{5}}
   & -\frac{s_2}{\sqrt{5}} & -\sqrt{\frac{2}{5}} s_2 &
   \frac{s_1}{\sqrt{5}} \\
 \sqrt{\frac{2}{5}} s_1 & 0 & \sqrt{\frac{2}{5}} s_2 & -\sqrt{\frac{2}{5}}
   s_2 & 0 & -\sqrt{\frac{2}{5}} s_1 \\
 \frac{s_2}{\sqrt{5}} & -\sqrt{\frac{2}{5}} s_2 & \frac{s_1}{\sqrt{5}}
   & \frac{s_1}{\sqrt{5}} & -\sqrt{\frac{2}{5}} s_1 &
   \frac{s_2}{\sqrt{5}} \\
\end{array}
\right) \nonumber \\
T &= \left(
\begin{array}{cccccc}
 e^{-\frac{7 i \pi }{120}} & 0 & 0 & 0 & 0 & 0 \\
 0 & e^{\frac{49 i \pi }{60}} & 0 & 0 & 0 & 0 \\
 0 & 0 & e^{\frac{17 i \pi }{120}} & 0 & 0 & 0 \\
 0 & 0 & 0 & e^{-\frac{103 i \pi }{120}} & 0 & 0 \\
 0 & 0 & 0 & 0 & e^{\frac{i \pi }{60}} & 0 \\
 0 & 0 & 0 & 0 & 0 & e^{\frac{113 i \pi }{120}} \\
\end{array}
\right)
\label{eq:tricritst}
\end{align}
where $s_1 = \sin{\left(\frac{2\pi}5\right)}$ and $s_2 = \sin{\left(\frac{4\pi}5\right)}$.

$\X^{(1;1)}$ is simply composed of the characters, $\chi_{(r,s)}$, of the six primary operators of $\mathcal{M}(5,4)$, which is given by (see e.g. \cite{DiFrancesco:1997nk})
\be
\chi_{(r,s)}(\tau) = K_{r,s}^{(5,4)}(\tau) - K_{r,-s}^{(5,4)}(\tau)
\ee
where
\be
K_{r,s}^{(p, p')}(\tau) = \frac{1}{\eta(\tau)}\sum_{n\in \mathbb{Z}}q^{\frac{2(p p' n + p r - p' s)^2}{4pp'}}
\ee
and $\eta(\tau)$ is the Dedekind eta function.
In our convention we then have
\begin{align}
\X^{(1;1)} = \begin{pmatrix} \chi_{(1,1)} \\ \chi_{(2,1)} \\ \chi_{(1,2)} \\ \chi_{(1,3)} \\ \chi_{(2,2)} \\ \chi_{(3,1)}\end{pmatrix}.
\end{align}
Now define the following operators:
\begin{align}
\nabla &= \frac{E_{10}(\tau)}{\eta(\tau)^{24}}\frac{d}{(2\pi i)d \tau} = \frac{E_{10}(\tau)}{\eta(\tau)^{24}} q\partial_q \nonumber \\
S_k &= \frac{12}{2\pi i} \frac{d}{d\tau} - k E_2(\tau) = 12 q \partial_q - k E_2(\tau) \nonumber\\
S^{(2)} &= \frac{E_8(\tau)}{\eta(\tau)^{24}} S_2 S_0 \nonumber \\
S^{(3)} &= \frac{E_6(\tau)}{\eta(\tau)^{24}} S_4 S_2 S_0 \nonumber \\
S^{(4)} &= \frac{E_4(\tau)}{\eta(\tau)^{24}} S_6 S_4 S_2 S_0 
\label{eq:differential_operator}
\end{align}
where $E_k(\tau)$ are the (quasi)-Eisenstein series
\begin{align}
E_2(\tau) &= 1-24\sum_{n=1}^{\infty}\frac{n q^n}{1-q^n} \nonumber \\
E_4(\tau) &= 1 + 240 \sum_{n=1}^{\infty} \frac{n^3 q^n}{1-q^n} \nonumber\\
E_6(\tau) &= 1-504\sum_{n=1}^{\infty} \frac{n^5 q^n}{1-q^n} \nonumber \\
E_8(\tau) &= E_4(\tau)^2 \nonumber \\
E_{10}(\tau) &= E_4(\tau) E_6(\tau).
\end{align}
We then have
\begin{align}
\X^{(2;1)} &= \frac{3165679}{94770} \X^{(1;1)} + \frac{32933}{44226} \nabla \X^{(1;1)} + \frac{2329}{2274480} J(\tau) \X^{(1;1)} \nonumber \\ &~~~~~~+ \frac{1975}{30618} S^{(2)} \X^{(1;1)} - \frac{50}{2187} S^{(3)} \X^{(1;1)} -\frac{1000}{199017} S^{(4)} \X^{(1;1)}  \nonumber \\
\X^{(3;1)} &= -\frac{7640171}{36450} \X^{(1;1)} + \frac{4265}{243} \nabla \X^{(1;1)} -\frac{46991}{874800} J(\tau) \X^{(1;1)}\nonumber \\ &~~~~~~ -\frac{4163}{8748} S^{(2)} \X^{(1;1)} + \frac{50}{2187} S^{(3)} \X^{(1;1)} +\frac{100}{2187} S^{(4)} \X^{(1;1)}\nonumber \\
\X^{(4;1)} &= -\frac{24177461}{947700} \X^{(1;1)}-\frac{983}{6318} \nabla \X^{(1;1)} -\frac{5831}{22744800} J(\tau) \X^{(1;1)}\nonumber \\ &~~~~~~ -\frac{71}{17496} S^{(2)} \X^{(1;1)} + \frac{25}{2187} S^{(3)} \X^{(1;1)} +\frac{50}{28431} S^{(4)} \X^{(1;1)} \nonumber \\
\X^{(5;1)} &=  \frac{14243257}{36450} \X^{(1;1)} -\frac{12799}{486} \nabla \X^{(1;1)} +\frac{798847}{874800} J(\tau) \X^{(1;1)}\nonumber \\ &~~~~~~ +\frac{2543}{4374} S^{(2)} \X^{(1;1)} + \frac{50}{2187} S^{(3)} \X^{(1;1)} -\frac{200}{2187}S^{(4)} \X^{(1;1)} \nonumber \\
\X^{(6;1)} &= -\frac{1}{306} \nabla \X^{(2;1)} - \frac{71}{36720} J(\tau) \X^{(2;1)} + \frac{2425}{153} \X^{(1;1)} + \frac12 \X^{(2;1)} \nonumber \\ &~~~~~~ -\frac{53}{765} \X^{(3;1)} + \frac{155}{51} \X^{(4;1)} - \frac{1}{85} \X^{(5;1)}.
\label{eq:tri-ising_chi}
\end{align}

\section{Strategies to compute the Bantay-Gannon characters}
The Bantay-Gannon construction provides us a method to construct partition functions with large $c$ and large gap: pick a complete basis of zero-gap forms $\X^{(\eta;1)}$ and use the recursion relation (\ref{eq:recursion-relation}) to raise both the gap and central charge. What the framework does not tell us, however, is how to find the complete basis $\X^{(\eta;1)}$. Since we are building the large $c$ models from minimal model characters, which is the $\X^{(1;1)}$ form of each basis, all the information of the basis must be encoded in $\X^{(1;1)}$, the matrix representation of $S$ and $T$, and the choice of $\Lambda$. Decoding this information turns out to be highly nontrivial. In this section, we describe some explicit strategies that we use to obtain the basis elements $\X^{(\eta;1)}$.

\subsection{Derivative operator method}
A straightforward way to compute the basis from the minimal model characters is to use the differential operators in (\ref{eq:differential_operator}). Each form $\X$ in $\CM (\rho)$ is determined by its polar part $\CP \X$, and a differential operator in  (\ref{eq:differential_operator}) maps the form to another form $\X^\prime$ in $\CM (\rho)$ with a different polar part $\CP \X^\prime$. One hopes with an infinite supply of differential operators such as $\nabla$, $S^{(2)}$, and $S^{(4)}$, the set $\mathbb C [J, \nabla, S^{(2)}, S^{(4)}, \cdots ]$ acting on $\X^{(1;1)}$ would span the whole module $\CM (\r)$, and one could find the specific linear combinations that give $\X^{(\eta;1)}$ by canceling out all other polar parts. Indeed this works at least for small $m$. For example, the $m=4$ tri-critical Ising model basis in (\ref{eq:tri-ising_chi}) is obtained this way. However at large $m$, the number of differential operators needed to span the basis grows rapidly and this method becomes inefficient. 

\subsection{Rademacher sum}
\label{sec:radsum}
For both zero weight and negative weight vector-valued modular forms, $\X_\n(q) = f_{\n}(n) q^{n+\Lambda_\n}$, the coefficients of positive power terms in the $q$-expansion can be obtained from a ``Rademacher sum'' (see e.g. \cite{Dijkgraaf:2000fq}),
\begin{align}
  f_\n(n) =& \, 2\pi \sum_{c=1}^{\infty} \sum_{\m=1}^{\text{dim} \X }
  c^{w-2} Kl(n,\n,m,\m,c) \sum_{m+\Lambda_\m<0} f_{\m}(m) \nn \\
  & \times \left( 2\pi\abs{m+\Lambda_\mu} \right)^{1-w} \tilde I_{1-w} \left( \frac{4\pi}{c} \sqrt{\abs{m+\Lambda_\m}(n+\Lambda_\n)} \right) ~,
\end{align}
where $w$ is the weight of the form, $Kl(n,\n,m,\m,c)$ is the generalized Kloosterman sum
\begin{align}
  Kl(n,\n,m,\m,c) &\equiv \sum_{0<-d<c;(d,c)=1} 
  e^{2\pi i (n+\Lambda_\n) \frac{d}{c}} \,
  M(\gamma_{c,d})_{\n\m}^{-1} \,
  e^{2\pi i (m+\Lambda_\m) \frac{a}{c}}~,
\end{align}
and $\tilde I$ is related to the Bessel $I$ function
\begin{align}
  \tilde I_\rho(z) &\equiv \left( \frac{z}{2} \right)^{-\rho} I_{\rho}(z)  ~.
\end{align}
For simplicity let us cover up all the details of the sum and express the operation of the Rademacher sum as
\begin{align}
  \X = \CF_w \[ \CP \X \] ~,
\end{align}
i.e. an operation that takes in some polar part $\CP \X$ and outputs the full vector-valued modular form $\X$. The operation is linear in its argument, the polar part.

\subsection{Negative weight Rademacher sum method}
A more efficient method than described above combines the Rademacher sum method with the Bantay-Gannon construction. 
Suppose we want to evaluate the character $\X^{(1;m)}$, which has only one polar term $q^{-m+\Lambda_1}$ at the first component. Consider a modified form $\widetilde\X(q) = \frac{1}{\Delta(q)}\X^{(1;m)} (q)$. The modified form has weight $-12$. The polar part of its first component is 
\begin{align}
  \CP \widetilde\X_1(q) = q^{\Lambda_1}\left(
     q^{-m-1}+24q^{-m}+324 q^{-m+1}+\cdots + x_1 q^{-1}
    \right)~,
\end{align}
with an unknown coefficient $x$.
For other components $\eta \neq 1$, the polar parts are $\CP \widetilde\X_\eta(q) = x_\eta q^{-1}$, with unknown coefficients $x_\eta$.
Thus all $q^{-1}$ coefficients of the modified form are unknown.
By going to $w=-12$ we take advantage that Rademacher sum of a negative weight modular form converges much faster than the $w=0$ case, but we have to deal with the unknown polar part. Since Rademacher sum is linear, we can always separate the contribution from these unknowns
\begin{align}
  \CP \widetilde\X 
  &= \left. \CP \widetilde\X \right|_{x=0} 
      + \sum_\eta x_\eta q^{-1+\Lambda_\eta} \,\mathbf e_\eta \\
  \CF_{w=-12}\[\CP \widetilde\X\] 
  &= \CF_{w=-12}\[\left. \CP \widetilde\X \right|_{x=0}\]
    + \sum_\eta x_\eta \CF_{w=-12}\[ q^{-1+\Lambda_\eta} \,\mathbf e_\eta  \]
    \label{eq:rad_linear_split}
\end{align} 
\begin{itemize}
  \item 
    The first term on the RHS of (\ref{eq:rad_linear_split}) ignores the unknown part and just feeds the known part $\left. \CP \widetilde\X_1 \right|_{x=0}$ into the Rademacher sum, and obtains the summed form as a $q$-expansion $F_{0}(q) \equiv \CF_{w=-12}\[\left. \CP \widetilde\X \right|_{x=0}\]$. Since the input is not the actual polar part of the form $\widetilde\X$, we expect that the result $F_{0}(q)$ is not quite the form $\widetilde\X$, but a new form as a linear combination dominated by $\widetilde\X$. We can read off the $q^{-1}$ coefficient of component $\xi$ of this new form as $F_0[-1]_\xi$. 
  \item 
    The second term on the RHS of (\ref{eq:rad_linear_split}) computes the contribution from each missing polar term individually. Define $F_{\eta}(q) \equiv \CF_{w=-12}\[q^{-1+\Lambda_\eta} \,\mathbf e_\eta \]$ as the form obtained from Rademacher sum of each individual $q^{-1}$ polar term in component $\eta$. Then we can read off the $q^{-1}$ coefficient of component $\xi$ of each new form $F_{\eta}(q)$ as $F_\eta[-1]_\xi$.
  \item
    The LHS of (\ref{eq:rad_linear_split}) should be the form $\widetilde\X$ itself, since the Rademacher sum is determined by its polar part.
\end{itemize}
Knowing how each term in (\ref{eq:rad_linear_split}) behaves, we can rewrite the equation as 
\begin{align}
  \widetilde\X(q) = F_{0}(q) + \sum_\eta x_\eta F_{\eta}(q)
\end{align} 
and take the $q^{-1}$ coefficient on both hands
\begin{align}
  x_\xi &= F_0[-1]_\xi + \sum_\eta x_\eta F_\eta[-1]_\xi \nn \\
  0 &= F_0[-1]_\xi + \sum_\eta  M_{\xi \eta} x_\eta.
  \label{eq:consistency_equation}
\end{align}
where $M_{\xi \eta} \equiv \left( F_\eta[-1]_\xi - \delta_{\xi\eta} \right)$.
Thus the unknown part of the modified form $\widetilde\X$ can be fixed by requiring that the Rademacher sum is consistent and linear. In practice, one can compute the coefficients $F_0[-1]_\xi$ and $F_\eta[-1]_\xi$ straightforwardly from the Rademacher sum, solve the linear consistency equation (\ref{eq:consistency_equation}), then use the updated polar part to obtain the correct Rademacher sum and obtain $\widetilde\X$, and finally put it back to the original form $\X^{(1;m)}(q) = \Delta(q) \widetilde\X(q)$. This method is equivalent to the $w=0$ Rademacher sum but converges much faster. 

It is not guaranteed that the linear equation (\ref{eq:consistency_equation}) is always determined. In practice, matrix $M_{\xi \eta}$ can be degenerate and the equation is underdetermined, for some choices of $\Lambda$. Then one need choose another $\Lambda$ consistent with (\ref{eq:consistent-lambda}), which may change what terms are considered to be polar and non-polar. For all the examples shown in this paper, it is possible to find such a $\Lambda$ that (\ref{eq:consistency_equation}) has a unique solution.
The strategy used explicitly in this project is
\begin{enumerate}
  \item Choose all elements of the $\Lambda$ matrix $\lambda_i$ so that $0\leq\lambda_i<1$. This will usually give a trace larger than the constraint (\ref{eq:consistent-lambda}). 
  \item While keeping the vacuum $\lambda_1$ unchanged, redefine the largest $\lambda_\xi$'s as $\lambda_\xi \rightarrow \lambda_\xi - 1$ until the trace of $\Lambda$ satisfies the constraint (\ref{eq:consistent-lambda}).
  \item Try evaluating the Rademacher sum and obtain the matrix 
  $M_{\xi \eta}$ 
  that appears in (\ref{eq:consistency_equation}). Compute the rank of the matrix.
  \item If matrix is full rank then (\ref{eq:consistency_equation}) has a unique solution. Otherwise, beginning from the lowest $\lambda_\xi$ and the highest $\lambda_\eta$, perform the ``exchange''
    \begin{align}
      \lambda_\xi &= \lambda_\xi + 1 \nn \\ 
      \lambda_\eta &= \lambda_\eta - 1 \nn~.
    \end{align}
  Meanwhile, the $\xi^{\text{th}}$ and $\eta^{\text{th}}$ rows and columns of $M_{\xi \eta}$ need to be computed again. If this step increases the rank of $M_{\xi \eta}$ then keep this choice of $\Lambda$. Otherwise discard the change. Keep performing this step until $M_{\xi \eta}$ has full rank.
\end{enumerate}
In practice, the pairs of $\Lambda$ that need to be exchanged seems to grow slowly with $m$. At $m=10$, no such exchange is needed. At $m=20$ one needs to exchange one pair of $\Lambda$ and at $m=37$ one needs to exchange 2 pairs of $\Lambda$. We do not know why some choices of $\Lambda$ (that satisfy (\ref{eq:consistent-lambda})) are better than others for determining the Bantay-Gannon characters. 

\section{Minimal model \texorpdfstring{$\sltz$}{Sl2Z} representations}
The unitary minimal model representations can be indexed by an integer $m$ with states labeled by two integers, $(r,s)$ that satisfy
\es{states}{
1\leq r \leq \frac{m}{2}, \ 1\leq s \leq m -1 &: m \, \textrm{even}\\
1\leq r \leq m, \ 1\leq s < \frac{m -1}{2} &: m \, \textrm{odd}\,,
}
or with the identification $(r,s)\equiv(m+1-r,m-s)$, 
\es{states_simp}{
	1\leq r \leq m, \ 1\leq s \leq m -1~.
}
These states transform under an $\frac{m(m-1)}2$ dimensional representations of $\sltz$ given by,
\es{mmsl2zrep}{
S_{(r,s)(r^{\prime},s^{\prime})}&=(-1)^{(r+s) (r^{\prime}+s^{\prime})}2 \sqrt{\frac{2}{m (m+1)}} \sin \left(\frac{\pi r r^{\prime}}{m+1}\right) \sin \left(\frac{\pi s s^{\prime}}{m}\right)\\
T_{(r,s)(r^{\prime},s^{\prime})}&=e^{2\pi i \left(h_{r,s}-\frac{c_{m}}{24}\right)}\delta_{(r,s)(r^{\prime},s^{\prime})}\,.
}
Here we have introduced the chiral dimension, $h_{r,s}$, and minimal model central charge, $c_{m}$,
\es{handc}{
h_{r,s} &= \frac{(m r-(m+1) s)^2-1}{4 m (m+1)} \\
c_{m}&=1-\frac{6}{m (m+1)}\,.
}

\section{Closed Form of Minimal Model Partition Functions}
\label{app:minmodelZ}

The characters of minimal models are known in closed-from (see for example Eqn (8.14)-(8.17) of \cite{DiFrancesco:1997nk}). A character labeled
$(r,s)$ from the index $m$ minimal model is given by
\begin{align}
  \X_\m(\tau) = \sum_n \frac{q^{E_{r,s,n}}-q^{E_{r,-s,n}}}{\eta(\tau)}
\end{align}
where $q^{E_{r,s,n}} \equiv e^{2i\pi \tau E_{r,s,n}}$ and
\begin{equation}
  \label{eq:singleQ}
  E_{r,s,n} \equiv \frac{(2 m (m+1) n+m r-(m+1) s)^2}{4 m (m+1)}.
\end{equation}
We choose the partition function to be diagonal, i.e. $ Z = \X_\mu
\bar\X^{\mu}$, where $\mu(r,s)$ is the label of the character. We
take the projection $\bar\tau = -\tau$ so the projected partition function is just sum of
each character squared. Finally we will define $Z_m^0(\tau):=2\eta(\tau)^2 Z(\tau,\bar\tau=-\tau)$. We then get
\begin{align}
  \label{eq:summand}
  Z_m^0(\tau) =& \sum_{r=1}^m \sum_{s=1}^{m-1} \sum_{n_1=-\infty}^{\infty}
      \sum_{n_2=-\infty}^{\infty}
      \left(e^{2 i \pi  \tau  E_{r,s,n_1}} -
      e^{2 i \pi  \tau E_{r,-s,n_1}}\right)
      \left(e^{2 i \pi  \tau E_{r,s,n_2}} -
      e^{2 i \pi  \tau E_{r,-s,n_2}}\right) \nn \\
    =& \sum_{r,s,n_1,n_2} \left(
      e^{2 i \pi  \tau  (E_{r,s,n_1} + E_{r,s,n_2})}
      -e^{2 i \pi  \tau  (E_{r,s,n_1} + E_{r,-s,n_2})} \right. \nn \\
      & \quad \quad \quad \quad \left. -e^{2 i \pi  \tau  (E_{r,-s,n_1} + E_{r,s,n_2})}
      +e^{2 i \pi  \tau  (E_{r,-s,n_1} + E_{r,-s,n_2})}
        \right) \nn \\
    =& \sum_{r,s,n_1,n_2} S_{r,s,n_1,n_2}^I - S_{r,s,n_1,n_2}^{II} - S_{r,s,n_1,n_2}^{III} + S_{r,s,n_1,n_2}^{IV}~.
\end{align}
In the last line we split the sum into four terms. It is convenient to group $I$
and $IV$ together, and $II$ and $III$ together and deal with each group
separately. The notation $S$ should not be confused with the modular transformation.

\textbf{I+IV}

In this subsection we do the following sum:
\begin{align}
  \label{eq:IandIV}
  \sum_{r,s,n_1,n_2} S_{r,s,n_1,n_2}^{I} + S_{r,s,n_1,n_2}^{IV}~.
\end{align}
It is convenient in this section to relabel $n_1$ and $n_2$ as
\begin{align}
  n_1 &:= \frac{n_a+n_b}{2} \nn \\
  n_2 &:= \frac{n_a-n_b}{2} \nn \\
  n_a &= n_b, n_b \pm 2, n_b \pm 4 \cdots \nn \\
  n_b &= 0, \pm 1, \pm 2 \cdots~.
\end{align}
So $n_a$ and $n_b$ have the same parity. Here we choose $n_a$ to
``jump'' with space 2, which leaves the space to map $S^{IV}$ as the
``missing'' terms in $S^{I}$. To see this, first observe the
relabelled sum has symmetries:
\begin{align}
  \label{eq:S14SymTranslation}
  S_{r,s \pm m,n_a,n_b}^{I,IV} &= S_{r,s,n_a \mp 1,n_b}^{I,IV} \\
  \label{eq:S14SymCross}
  S_{r,m-s,n_a,n_b}^{IV} &= S_{r,s-m,n_a,n_b}^{I} \\
  \label{eq:S14SymFactor}
  S_{r,s,n_a,n_b}^{I} &= S_{n_b}^{I,(0)}S_{r,s,n_a,n_b=0}^{I}~,
\end{align}
where
\begin{align}
  \label{eq:S14Const}
  S_{n_b}^{I,(0)} = e^{i \pi  m (m+1) n_b^2 \tau}~.
\end{align}
First we use (\ref{eq:S14SymCross}) to map $S^{IV}$ as the
negative $s$ terms in $S^{I}$. Then (\ref{eq:S14SymTranslation})
allows us to interpret $n_a$ as extensions of $s$ beyond its
domain. Finally we factor out the piece only dependent on $n_b$
according to (\ref{eq:S14SymFactor}). The procedures are as follows:
\begin{align}
  &\sum_{r=0}^m \sum_{s=0}^{m-1} \sum_{n_a=n_b-2\infty}^{n_b+2\infty}
      \sum_{n_b=-\infty}^{\infty} S_{r,s,n_a,n_b}^{I} +
    S_{r,s,n_a,n_b}^{IV} \nn \\
  =& \sum_{r=1}^m \sum_{s=-m+1}^{m-1} \sum_{n_a=n_b-2\infty}^{n_b+2\infty}
     \sum_{n_b=-\infty}^{\infty} S_{r,s,n_a,n_b}^{I} \nn \\
  =& \sum_{r=1}^m \sum_{s=-\infty}^{\infty}
     \sum_{n_b=-\infty}^{\infty} S_{r,s,n_a=0,n_b}^{I} \nn \\
  =& \left( \sum_{n_b=-\infty}^{\infty} S_{n_b}^{I,(0)} \right)
     \left( \sum_{r=1}^m \sum_{s=-\infty}^{\infty} S_{r,s,n_a=0,n_b=0}^{I}\right)~.
\end{align}
Note that in the above summation $r=0$ and $s=0$ are trivially
included in the sum since their contribution cancel out.
The $n_b$ piece in (\ref{eq:S14Const}) sums to a $\vartheta _3$
\begin{align}
  \sum_{n_b=-\infty}^{\infty} S_{n_b}^{I,(0)} =
  \sum_{n_b=-\infty}^{\infty} e^{i \pi  m (m+1) n_b^2 \tau} =
  \vartheta _3\left(0,e^{i \pi  m (m+1) \tau }\right)~.
\end{align}
We can use Poisson resummation to simplify the infinite sum over $s$
\begin{align}
  \label{eq:S14SumPs}
  \sum_{s=-\infty}^{\infty} S_{r,s}^{I} =
  \sum_{p_s=-\infty}^{\infty}\widehat{S}_{r,p_s}^{I}
  =\sum_{p_s=-\infty}^{\infty} \frac{e^{-\frac{i \pi  m p_s (p_s-2 r \tau )}{(m+1)
  \tau }}}{\sqrt{-\frac{i (m+1)\tau }{m}}}~,
\end{align}
where the convention is chosen as
\begin{align}
  \widehat{f}(p_s) :=  \int f(s) e^{2\pi i s p_s} ds ~.
\end{align}
Since (\ref{eq:S14SumPs}) is periodic in $r$, the geometric sum of
$r$ over a period vanishes unless the ratio is 1
\begin{align}
  \sum_{r=0}^{m} \frac{e^{-\frac{i \pi  m p_s (p_s-2 r \tau )}{(m+1)
  \tau }}}{\sqrt{-\frac{i (m+1)\tau }{m}}} = 0
  ~~~{\rm unless}~~~ \frac{m \, p_s}{m+1} = \mathbb{Z}~.
\end{align}
Since $m$ and $m+1$ are always coprime, $p_s = (m+1)p$, $p \in \mathbb{Z}$ and
sum over $r$ is trivial. The $p$ sum results in another $\vartheta _3$
\begin{align}
  \label{eq:S14Rest}
  \sum_{p_s=-\infty}^{\infty} \sum_{r=0}^{m} \widehat{S}_{r,p_s}^{I}
  = \sum_{p=-\infty}^{\infty} \frac{(m+1) e^{-\frac{i \pi  m (m+1) p^2}{\tau }}}
  {\sqrt{-\frac{i (m+1) \tau }{m}}}
  =\frac{(m+1) \vartheta _3\left(
    0,e^{-\frac{i m (m+1) \pi }{\tau}}
  \right)}
  {\sqrt{-\frac{i (m+1)\tau }{m}}}~.
\end{align}
Combine (\ref{eq:S14Const}) and (\ref{eq:S14Rest}), we arrive at a
closed form of sum $I$ and $IV$
\begin{align}
  \label{eq:S14Final}
  \sum_{r,s,n_1,n_2} S_{r,s,n_1,n_2}^{I} + S_{r,s,n_1,n_2}^{IV} =
  \frac{(m+1) \vartheta _3\left(
  0,e^{-\frac{i m (m+1) \pi }{\tau}}\right)
  \vartheta_3\left(
  0,e^{i m (m+1) \pi  \tau }\right)}
  {\sqrt{-\frac{i (m+1) \tau }{m}}}~.
\end{align}

\textbf{II+III}

\label{sec:IIandIII}
In this subsection we do the $II$ and $III$ sum. Similar to the last
subsection, $S^{II,III}$ also have symmetries:
\begin{align}
  S_{r \pm (m+1),s \pm m,n_1,n_2}^{II} &= S_{r,s,n_1 \mp 1,n_2}^{II} \\
  S_{r \pm (m+1),s \mp m,n_1,n_2}^{II} &= S_{r,s,n_1,n_2 \pm 1}^{II} \\
  S_{r,s,n_1,n_2}^{III} &= S_{r,-s,n_1,n_2}^{II}~.
\end{align}
These symmetries help us reinterpret $n_1$ and $n_2$ as extensions of
$r$ and $s$ beyond their domain:
\begin{align}
  \sum_{r=1}^m \sum_{s=1}^{m-1} \sum_{n_1=-\infty}^{\infty}
  \sum_{n_2=-\infty}^{\infty} S_{r,s,n_1,n_2}^{II} +
  S_{r,s,n_1,n_2}^{III} =
  \sum_{r=-\infty}^{\infty} \sum_{s=-\infty}^{\infty} S_{r,s,n_1=0,n_2=0}^{II}~.
\end{align}
Since both $r$ and $s$ are unbounded we can perform Poisson
resummation on both indices, resulting in $\vartheta _3$
\begin{align}
  \label{eq:S23Final}
  &\sum_{r=-\infty}^{\infty} \sum_{s=-\infty}^{\infty}
    S_{r,s}^{II} =
    \sum_{p_r=-\infty}^{\infty} \sum_{p_s=-\infty}^{\infty}
    \widehat{S}_{p_r,p_s}^{II} =
    \sum_{p_s,p_r}
    -\frac{\exp \left(
    -\frac{i \pi  \left(m^2 p_s^2+(m+1)^2 p_r^2\right)}
    {m (m+1) \tau}\right)}
    {\sqrt{-\frac{i m \tau }{m+1}} \sqrt{-\frac{i (m+1) \tau }{m}}}
    \nn \\
  =&-\frac{1}{\tau ^2}
     \sqrt{-\frac{i m \tau }{m+1}} \sqrt{-\frac{i (m+1) \tau }{m}}
     \ \times
     \left\{
     -\vartheta _3\left(0,e^{-\frac{i m \pi }{(m+1) \tau }}\right)
     \vartheta _3\left(0,e^{-\frac{i(m+1) \pi }{m \tau }}\right)
     \right. \nn \\
  &\quad \left.
     -\vartheta _3\left(0,e^{-\frac{i (m+1) \pi }{m \tau }}\right)
    \right\}
    + \frac{
    \vartheta _3\left(0,
    e^{-\frac{i (m+1)^2 \pi }{\left(m^2+m\right)\tau}}
    \right)}
    {\sqrt{-\frac{i m \tau }{m+1}} \sqrt{-\frac{i (m+1) \tau }{m}}}~.
\end{align}

\textbf{Putting it all together}

Substitute (\ref{eq:S14Final}) and (\ref{eq:S23Final}) in
(\ref{eq:summand}) we obtain the closed form of the whole sum
\begin{align}
  Z_m^0(\tau) = &\frac{1}{\tau ^2}\sqrt{-\frac{i (m+1) \tau }{m}}
  \Bigg\{
  i m \tau  \vartheta _3\left(0,e^{-\frac{i m (m+1) \pi }{\tau}}\right)
  \vartheta _3\left(0,e^{im (m+1) \pi  \tau }\right) \nn \\
  &\quad+ \sqrt{-\frac{i m \tau }{m+1}}
  \vartheta _3\left(0,e^{-\frac{i (m+1) \pi }{m \tau }}\right)
  \vartheta _3\left(0,e^{-\frac{i m \pi }{m \tau +\tau }}\right)
  \Bigg\}~.
\end{align}
Using $\vartheta_3(0,e^{i\pi\frac{-1}{\tau}}) =
\sqrt{-i \tau} \, \vartheta_3(0,e^{i\pi \tau})$ we can further
simplify it:
\begin{align}
  Z_m^0(\tau) = \vartheta _3\left(0,e^{\frac{i \pi  \tau }{m (m+1)}}\right)
  \vartheta _3\left(0,e^{i m (m+1) \pi \tau }\right)
  -\vartheta _3\left(0,e^{\frac{i m \pi  \tau }{m+1}}\right)
  \vartheta_3\left(0,e^{\frac{i (m+1) \pi  \tau }{m}}\right)~.
\end{align}
Finally we put back the Dedekind eta function and a factor of
$\frac{1}{2}$ from double-counting $(r,s) \equiv (m+1-r,m-s)$. The projected
partition function of the $m^{\text{th}}$ minimal model is
\begin{align}
  Z_m(\tau) = \frac{Z_m^0(\tau)}{2\eta(\tau)^2}~.
\end{align}

\section{Spectral form factor size computations}
\subsection{Vacuum-vacuum contribution}\label{app:vv}
\textbf{Late time $m$ scaling}:
In this subsection we derive the asymptotic behavior of the vacuum-vacuum contribution to the SFF quoted in \ref{vvatlate}. 
The integer averaged SFF is given by,
\es{vvatlateexact}{
\lim_{n\rightarrow\infty}\frac{1}{n}\left|\sum_{n^{\prime}=0}^{n}V\times V\right|^{2}&=\lim_{n\rightarrow\infty}\frac{1}{n}\left|\sum_{n^{\prime}=0}^{n}S_{0\alpha}T_{\alpha\beta}^{2n^{\prime}}S_{\beta 0}\right|^{2}\\
&=\sum_{\alpha}S_{0\alpha}^{4} + \sum_{\alpha,\beta : \,|E_{\alpha}-E_{\beta}| \in \mathbb{Z}^{+}}\left|S_{0\alpha}S_{0\beta}\right |^{2}
}

The quartic term is easy to estimate from equation \eqref{mmsl2zrep}.
\es{S4}{
\sum_{\alpha}S_{0\alpha}^{4}&=\frac{64}{m^{4}}\sum_{(r,s)}\sin^{4} \left(\frac{\pi r}{m+1}\right) \sin^{4} \left(\frac{\pi s}{m}\right)+\ldots\\
&\leq \frac{32}{m^{2}}+\ldots
}
The second term takes a bit more care, we have to estimate the number of pairs of states, whose dimension differ by an integer. 
The condition $h_{r,s} - h_{r^{\prime}, s^{\prime}}\in \mathbb{Z}$ can be massaged into the form:
\es{integerdiff}{
\frac{m (r-r^{\prime}) (r+r^{\prime})}{4 (m+1)}+\frac{(m+1) (s-s^{\prime}) (s+s^{\prime})}{4 m}+\frac{r^{\prime}s^{\prime}-r s}{2}\in\mathbb{Z}\,.
}
The number of solutions to this equation scales as $m^{2}$ (essentially for generic $m$, we need $r+r'=m+1$, $s+s' = m$, which gives us $\mathcal{O}(m^2)$ choices for $r, r', s, s'$), and so we can bound the entire vacuum-vacuum contribution at large $m$ as scaling like $1/m^{2}$.

\textbf{Finite time $m$ scaling:}

We can also bound the size of the vacuum-vacuum contribution at finite times.
\es{vvearly}{
S_{0\alpha}T_{\alpha\beta}^{2n}S_{\beta 0}&=\sum_{\alpha}S_{0\alpha}^{2}e^{4\pi i\phi_{\alpha}n}\\
&=\frac{2}{m(m+1)}\sum_{(r^{\prime},s^{\prime})}e^{4\pi i\phi_{\alpha}n}\sin^{2}\left(\frac{\pi r^{\prime}}{m+1}\right)\sin^{2}\left(\frac{\pi s^{\prime}}{m}\right)\\
&\sim\frac{2e^{-4\pi i (1+\frac{c_{m}}{24})}}{m(m+1)}\int dr^{\prime}ds^{\prime} e^{\pi i nm(m+1)\left(\frac{r^{\prime}}{m+1}-\frac{s^{\prime}}{m}\right)^{2}}\sin^{2}\left(\frac{\pi r^{\prime}}{m+1}\right)\sin^{2}\left(\frac{\pi s^{\prime}}{m}\right)\\
&\sim2e^{-4\pi i (1+\frac{c_{m}}{24})}\int dx dy e^{\pi i nm^{2}(x-y)^{2}}\sin^{2}\left(x\right)\sin^{2}\left(y\right)\\
&\sim\frac{C}{m\sqrt{n}}
}
\subsection{Heavy-heavy contribution}
We can also bound the heavy-heavy contribution to the normalized SFF at all times. 
\es{hhcont}{
H\times H&=S_{i\alpha}T_{\alpha\beta}^{2n}S_{\beta j}\hat{\chi}^{i}\hat{\chi}^{j}\\
&=\sum_{\alpha}v_{\alpha}^{2}e^{4\pi i \phi_{\alpha}n}\, \leq \, \sum_{\alpha}v_{\alpha}^{2}\\
&=\hat{\chi}^{i}\hat{\chi}_{i}\,\rightarrow\, C e^{-\frac{8\pi^{2}}{\beta}\Delta}
}
Here, we have defined the vector,
\es{vdef}{
v_{\alpha}&=S_{\alpha i}\hat{\chi}^{i}\,,
}
and the arrow in the last line of \eqref{hhcont} indicates the high temperature limit. 
\end{appendices}

%\newpage
\label{refs}

\bibliographystyle{utphys}
\bibliography{refs}

%bibliography generated by nb.bst v1.01 (C) 2003-2010 Niklas Beisert
\begin{thebibliography}{10}
\ifx\href\asklfhas\newcommand{\href}[2]{#2}\fi
\ifx\arxivref\asklfhas\newcommand{\arxivref}[2]{\href{http://arxiv.org/abs/#1}{#2}}\fi
\ifx\doiref\asklfhas\newcommand{\doiref}[2]{\href{http://dx.doi.org/#1}{#2}}\fi
\parskip 0pt
\normalsize

\bibitem{Maldacena:1997re}
J.~M. Maldacena,
\textit{``{The large N limit of superconformal field theories and
  supergravity}''},
Adv.~Theor.~Math.~Phys. \textbf{2}, 231 (1998),
\normalsize{\texttt{\arxivref{hep-th/9711200}{hep-th/9711200}}}.
%%CITATION = HEP-TH/9711200;%%

\bibitem{Heemskerk:2009pn}
I.~Heemskerk, J.~Penedones, J.~Polchinski \& J.~Sully,
\textit{``{Holography from Conformal Field Theory}''},
\doiref{10.1088/1126-6708/2009/10/079}{JHEP \textbf{0910}, 079 (2009)},
\normalsize{\texttt{\arxivref{0907.0151}{arXiv:0907.0151}}}.
%%CITATION = ARXIV:0907.0151;%%

\bibitem{Hartman:2014oaa}
T.~Hartman, C.~A. Keller \& B.~Stoica,
\textit{``{Universal Spectrum of 2d Conformal Field Theory in the Large c
  Limit}''},
\doiref{10.1007/JHEP09(2014)118}{JHEP \textbf{1409}, 118 (2014)},
\normalsize{\texttt{\arxivref{1405.5137}{arXiv:1405.5137}}}.
%%CITATION = ARXIV:1405.5137;%%

\bibitem{Witten:2007kt}
E.~Witten,
\textit{``{Three-Dimensional Gravity Revisited}''},
\normalsize{\texttt{\arxivref{0706.3359}{arXiv:0706.3359}}}.
%%CITATION = ARXIV:0706.3359;%%

\bibitem{FLM}
I.~Frenkel, J.~Lepowsky \& A.~Meurman,
\textit{``Vertex Operator Algebras and the Monster''},
Elsevier Science (1989).

\bibitem{Gaberdiel:2007ve}
M.~R. Gaberdiel,
\textit{``{Constraints on extremal self-dual CFTs}''},
\doiref{10.1088/1126-6708/2007/11/087}{JHEP \textbf{0711}, 087 (2007)},
\normalsize{\texttt{\arxivref{0707.4073}{arXiv:0707.4073}}}.
%%CITATION = ARXIV:0707.4073;%%

\bibitem{Gaiotto:2008jt}
D.~Gaiotto,
\textit{``{Monster symmetry and Extremal CFTs}''},
\doiref{10.1007/JHEP11(2012)149}{JHEP \textbf{1211}, 149 (2012)},
\normalsize{\texttt{\arxivref{0801.0988}{arXiv:0801.0988}}}.
%%CITATION = ARXIV:0801.0988;%%

\bibitem{Gaberdiel:2008pr}
M.~R. Gaberdiel \& C.~A. Keller,
\textit{``{Modular differential equations and null vectors}''},
\doiref{10.1088/1126-6708/2008/09/079}{JHEP \textbf{0809}, 079 (2008)},
\normalsize{\texttt{\arxivref{0804.0489}{arXiv:0804.0489}}}.
%%CITATION = ARXIV:0804.0489;%%

\bibitem{Gaberdiel:2008xb}
M.~R. Gaberdiel, S.~Gukov, C.~A. Keller, G.~W. Moore \& H.~Ooguri,
\textit{``{Extremal N=(2,2) 2D Conformal Field Theories and Constraints of
  Modularity}''},
\doiref{10.4310/CNTP.2008.v2.n4.a3}{Commun.~Num.~Theor.~Phys. \textbf{2}, 743
  (2008)},
\normalsize{\texttt{\arxivref{0805.4216}{arXiv:0805.4216}}}.
%%CITATION = ARXIV:0805.4216;%%

\bibitem{Benjamin:2016aww}
N.~Benjamin, E.~Dyer, A.~L. Fitzpatrick, A.~Maloney \& E.~Perlmutter,
\textit{``{Small Black Holes and Near-Extremal CFTs}''},
\doiref{10.1007/JHEP08(2016)023}{JHEP \textbf{1608}, 023 (2016)},
\normalsize{\texttt{\arxivref{1603.08524}{arXiv:1603.08524}}}.
%%CITATION = ARXIV:1603.08524;%%

\bibitem{Shenker:2013pqa}
S.~H. Shenker \& D.~Stanford,
\textit{``{Black holes and the butterfly effect}''},
\doiref{10.1007/JHEP03(2014)067}{JHEP \textbf{1403}, 067 (2014)},
\normalsize{\texttt{\arxivref{1306.0622}{arXiv:1306.0622}}}.
%%CITATION = ARXIV:1306.0622;%%

\bibitem{Shenker:2013yza}
S.~H. Shenker \& D.~Stanford,
\textit{``{Multiple Shocks}''},
\doiref{10.1007/JHEP12(2014)046}{JHEP \textbf{1412}, 046 (2014)},
\normalsize{\texttt{\arxivref{1312.3296}{arXiv:1312.3296}}}.
%%CITATION = ARXIV:1312.3296;%%

\bibitem{Maldacena:2015waa}
J.~Maldacena, S.~H. Shenker \& D.~Stanford,
\textit{``{A bound on chaos}''},
\doiref{10.1007/JHEP08(2016)106}{JHEP \textbf{1608}, 106 (2016)},
\normalsize{\texttt{\arxivref{1503.01409}{arXiv:1503.01409}}}.
%%CITATION = ARXIV:1503.01409;%%

\bibitem{Cotler:2016fpe}
J.~S. Cotler, G.~Gur-Ari, M.~Hanada, J.~Polchinski, P.~Saad, S.~H. Shenker,
  D.~Stanford, A.~Streicher \& M.~Tezuka,
\textit{``{Black Holes and Random Matrices}''},
\doiref{10.1007/JHEP05(2017)118}{JHEP \textbf{1705}, 118 (2017)},
\normalsize{\texttt{\arxivref{1611.04650}{arXiv:1611.04650}}}.
%%CITATION = ARXIV:1611.04650;%%

\bibitem{Hellerman:2009bu}
S.~Hellerman,
\textit{``{A Universal Inequality for CFT and Quantum Gravity}''},
\doiref{10.1007/JHEP08(2011)130}{JHEP \textbf{1108}, 130 (2011)},
\normalsize{\texttt{\arxivref{0902.2790}{arXiv:0902.2790}}}.
%%CITATION = ARXIV:0902.2790;%%

\bibitem{Friedan:2013cba}
D.~Friedan \& C.~A. Keller,
\textit{``{Constraints on 2d CFT partition functions}''},
\doiref{10.1007/JHEP10(2013)180}{JHEP \textbf{1310}, 180 (2013)},
\normalsize{\texttt{\arxivref{1307.6562}{arXiv:1307.6562}}}.
%%CITATION = ARXIV:1307.6562;%%

\bibitem{Collier:2016cls}
S.~Collier, Y.-H. Lin \& X.~Yin,
\textit{``{Modular Bootstrap Revisited}''},
\doiref{10.1007/JHEP09(2018)061}{JHEP \textbf{1809}, 061 (2018)},
\normalsize{\texttt{\arxivref{1608.06241}{arXiv:1608.06241}}}.
%%CITATION = ARXIV:1608.06241;%%

\bibitem{Bae:2017kcl}
J.-B. Bae, S.~Lee \& J.~Song,
\textit{``{Modular Constraints on Conformal Field Theories with Currents}''},
\doiref{10.1007/JHEP12(2017)045}{JHEP \textbf{1712}, 045 (2017)},
\normalsize{\texttt{\arxivref{1708.08815}{arXiv:1708.08815}}}.
%%CITATION = ARXIV:1708.08815;%%

\bibitem{Dyer:2017rul}
E.~Dyer, A.~L. Fitzpatrick \& Y.~Xin,
\textit{``{Constraints on Flavored 2d CFT Partition Functions}''},
\doiref{10.1007/JHEP02(2018)148}{JHEP \textbf{1802}, 148 (2018)},
\normalsize{\texttt{\arxivref{1709.01533}{arXiv:1709.01533}}}.
%%CITATION = ARXIV:1709.01533;%%

\bibitem{Bantay:2005vk}
P.~Bantay \& T.~Gannon,
\textit{``{Conformal characters and the modular representation}''},
\doiref{10.1088/1126-6708/2006/02/005}{JHEP \textbf{0602}, 005 (2006)},
\normalsize{\texttt{\arxivref{hep-th/0512011}{hep-th/0512011}}}.
%%CITATION = HEP-TH/0512011;%%

\bibitem{Balasubramanian:2016ids}
V.~Balasubramanian, B.~Craps, B.~Czech \& G.~Sárosi,
\textit{``{Echoes of chaos from string theory black holes}''},
\doiref{10.1007/JHEP03(2017)154}{JHEP \textbf{1703}, 154 (2017)},
\normalsize{\texttt{\arxivref{1612.04334}{arXiv:1612.04334}}}.
%%CITATION = ARXIV:1612.04334;%%

\bibitem{Dyer:2016pou}
E.~Dyer \& G.~Gur-Ari,
\textit{``{2D CFT Partition Functions at Late Times}''},
\doiref{10.1007/JHEP08(2017)075}{JHEP \textbf{1708}, 075 (2017)},
\normalsize{\texttt{\arxivref{1611.04592}{arXiv:1611.04592}}}.
%%CITATION = ARXIV:1611.04592;%%

\bibitem{Maloney:2007ud}
A.~Maloney \& E.~Witten,
\textit{``{Quantum Gravity Partition Functions in Three Dimensions}''},
\doiref{10.1007/JHEP02(2010)029}{JHEP \textbf{1002}, 029 (2010)},
\normalsize{\texttt{\arxivref{0712.0155}{arXiv:0712.0155}}}.
%%CITATION = ARXIV:0712.0155;%%

\bibitem{Cardy:2014rqa}
J.~Cardy,
\textit{``{Thermalization and Revivals after a Quantum Quench in Conformal
  Field Theory}''},
\doiref{10.1103/PhysRevLett.112.220401}{Phys.~Rev.~Lett. \textbf{112}, 220401
  (2014)},
\normalsize{\texttt{\arxivref{1403.3040}{arXiv:1403.3040}}}.
%%CITATION = ARXIV:1403.3040;%%

\bibitem{DAlessio:2016rwt}
L.~D'Alessio, Y.~Kafri, A.~Polkovnikov \& M.~Rigol,
\textit{``{From quantum chaos and eigenstate thermalization to statistical
  mechanics and thermodynamics}''},
\doiref{10.1080/00018732.2016.1198134}{Adv.~Phys. \textbf{65}, 239 (2016)},
\normalsize{\texttt{\arxivref{1509.06411}{arXiv:1509.06411}}}.
%%CITATION = ARXIV:1509.06411;%%

\bibitem{Bantay:2007zz}
P.~Bantay \& T.~Gannon,
\textit{``{Vector-valued modular functions for the modular group and the
  hypergeometric equation}''},
\doiref{10.4310/CNTP.2007.v1.n4.a2}{Commun.~Num.~Theor.~Phys. \textbf{1}, 651
  (2007)},
\normalsize{\texttt{\arxivref{0705.2467}{arXiv:0705.2467}}}.
%%CITATION = ARXIV:0705.2467;%%

\bibitem{Bantay:2011}
P.~Bantay,
\textit{``{The Dimension of Spaces of Vector-Valued Modular Forms of Integer
  Weight}''},
\doiref{10.1007/s11005-013-0641-6}{Lett.~Math.~Phys. \textbf{103}, 1243
  (2013)},
\normalsize{\texttt{\arxivref{1104.1278}{arXiv:1104.1278}}}.
%%CITATION = ARXIV:1104.1278;%%

\bibitem{Dijkgraaf:2000fq}
R.~Dijkgraaf, J.~M. Maldacena, G.~W. Moore \& E.~P. Verlinde,
\textit{``{A Black hole Farey tail}''},
\normalsize{\texttt{\arxivref{hep-th/0005003}{hep-th/0005003}}}.
%%CITATION = HEP-TH/0005003;%%

\bibitem{Keller:2014xba}
C.~A. Keller \& A.~Maloney,
\textit{``{Poincare Series, 3D Gravity and CFT Spectroscopy}''},
\doiref{10.1007/JHEP02(2015)080}{JHEP \textbf{1502}, 080 (2015)},
\normalsize{\texttt{\arxivref{1407.6008}{arXiv:1407.6008}}}.
%%CITATION = ARXIV:1407.6008;%%

\bibitem{sachdev1993gapless}
S.~Sachdev \& J.~Ye,
\textit{``Gapless spin-fluid ground state in a random quantum Heisenberg
  magnet''},
Physical~review~letters \textbf{70}, 3339 (1993).

\bibitem{Kitaev:2017awl}
A.~Kitaev \& S.~J. Suh,
\textit{``{The soft mode in the Sachdev-Ye-Kitaev model and its gravity
  dual}''},
\doiref{10.1007/JHEP05(2018)183}{JHEP \textbf{1805}, 183 (2018)},
\normalsize{\texttt{\arxivref{1711.08467}{arXiv:1711.08467}}}.
%%CITATION = ARXIV:1711.08467;%%

\bibitem{Chen:2017yze}
H.~Chen, C.~Hussong, J.~Kaplan \& D.~Li,
\textit{``{A Numerical Approach to Virasoro Blocks and the Information
  Paradox}''},
\doiref{10.1007/JHEP09(2017)102}{JHEP \textbf{1709}, 102 (2017)},
\normalsize{\texttt{\arxivref{1703.09727}{arXiv:1703.09727}}}.
%%CITATION = ARXIV:1703.09727;%%

\bibitem{Esterlis:2016psv}
I.~Esterlis, A.~L. Fitzpatrick \& D.~Ramirez,
\textit{``{Closure of the Operator Product Expansion in the Non-Unitary
  Bootstrap}''},
\doiref{10.1007/JHEP11(2016)030}{JHEP \textbf{1611}, 030 (2016)},
\normalsize{\texttt{\arxivref{1606.07458}{arXiv:1606.07458}}}.
%%CITATION = ARXIV:1606.07458;%%

\bibitem{Harvey:2018rdc}
J.~A. Harvey \& Y.~Wu,
\textit{``{Hecke Relations in Rational Conformal Field Theory}''},
\doiref{10.1007/JHEP09(2018)032}{JHEP \textbf{1809}, 032 (2018)},
\normalsize{\texttt{\arxivref{1804.06860}{arXiv:1804.06860}}}.
%%CITATION = ARXIV:1804.06860;%%

\bibitem{Aharony:2018bad}
O.~Aharony, S.~Datta, A.~Giveon, Y.~Jiang \& D.~Kutasov,
\textit{``{Modular invariance and uniqueness of $T\bar{T}$ deformed CFT}''},
\normalsize{\texttt{\arxivref{1808.02492}{arXiv:1808.02492}}}.
%%CITATION = ARXIV:1808.02492;%%

\bibitem{DiFrancesco:1997nk}
P.~Di~Francesco, P.~Mathieu \& D.~Senechal,
\textit{``{Conformal Field Theory}''},
Springer-Verlag (1997),
New York.

\end{thebibliography}

\end{document}